\documentclass[reprint, aps,superscriptaddress,amsmath,amssym,prb]
{revtex4-2}

\usepackage{bm}
\usepackage{float}
\usepackage{graphicx}
\usepackage[usenames,dvipsnames]{color}
\usepackage[normalem]{ulem}
\usepackage{bm}
\usepackage{multirow}
\usepackage{titlesec}
\usepackage{ragged2e}
\usepackage[english]{babel}
\usepackage[T1]{fontenc}
\usepackage[table]{xcolor}
\usepackage[utf8]{inputenc}
\usepackage{lmodern}
\usepackage{babel}
\usepackage{amsmath}
\usepackage{amssymb}
\usepackage{envmath}
\usepackage{makecell}
\usepackage{comment}
\usepackage{geometry}
\usepackage{caption}
\usepackage{subcaption}
\usepackage{stackengine}
\usepackage{graphicx}
\usepackage{tikz}
\graphicspath{{images/}}
\usepackage{array}
\newcolumntype{P}[1]{>{\centering\arraybackslash}p{#1}}
\usepackage{parskip}

\usepackage{wrapfig}

\begin{document}
\title{Magnetic memory and distinct spin populations in ferromagnetic Co$_{3}$Sn$_{2}$S$_{2}$}
\author{Charles Menil}
\affiliation{Laboratoire de Physique et d'\'Etude des Mat\'eriaux \\ (ESPCI - CNRS - Sorbonne Universit\'e)\\ Universit\'e Paris Sciences et Lettres, 75005 Paris, France}
\author{Brigitte Leridon}
\affiliation{Laboratoire de Physique et d'\'Etude des Mat\'eriaux \\ (ESPCI - CNRS - Sorbonne Universit\'e)\\ Universit\'e Paris Sciences et Lettres, 75005 Paris, France}
\author{Antonella Cavanna}
\affiliation{Centre de Nanosciences et de Nanotechnologies, CNRS, Université Paris-Saclay, 91120 Palaiseau, France}
\author{Ulf Gennser}
\affiliation{Centre de Nanosciences et de Nanotechnologies, CNRS, Université Paris-Saclay, 91120 Palaiseau, France}
\author{Dominique Mailly}
\affiliation{Centre de Nanosciences et de Nanotechnologies, CNRS, Université Paris-Saclay, 91120 Palaiseau, France}

\author{Linchao Ding}
\affiliation{Wuhan National High Magnetic Field Center and School of Physics\\ Huazhong University of Science and Technology,  Wuhan  430074, China}
\author{Xiaokang Li}
\affiliation{Wuhan National High Magnetic Field Center and School of Physics\\ Huazhong University of Science and Technology,  Wuhan  430074, China}
\author{Zengwei Zhu}
\affiliation{Wuhan National High Magnetic Field Center and School of Physics\\ Huazhong University of Science and Technology,  Wuhan  430074, China}
\author{Beno\^it Fauqu\'e}
\affiliation{JEIP  (USR 3573 CNRS), Coll\`ege de France \\ Universit\'e Paris Sciences et Lettres, 75005 Paris, France}
\author{Kamran Behnia}
\affiliation{Laboratoire de Physique et d'\'Etude des Mat\'eriaux \\ (ESPCI - CNRS - Sorbonne Universit\'e)\\ Universit\'e Paris Sciences et Lettres, 75005 Paris, France}

\date{\today}

\begin{abstract}
Co$_{3}$Sn$_{2}$S$_{2}$, a ferromagnetic Weyl semi-metal with  Co atoms on a kagome lattice, has generated much recent attention. Experiments have identified a temperature scale below the Curie temperature. Here, we find that this magnet keeps a memory, when not exposed to a magnetic field sufficiently large to erase it.  We identify the driver of this memory effect as a small secondary population of spins, whose coercive field is significantly larger than that of the majority spins. The shape of the magnetization hysteresis curve has a threshold magnetic field set by the demagnetizing factor. These two field scales set the hitherto unidentified temperature scale, which is not a thermodynamic phase transition, but a crossing point between meta-stable boundaries. Global magnetization is well-defined, even when it is non-uniform, but  drastic variations in local magnetization point to a coarse energy landscape,  with the thermodynamic limit not achieved at micrometer length scales.

\end{abstract}
\maketitle
\section{Introduction}
\begin{figure}
    \centering
    \includegraphics[width=0.95\linewidth]{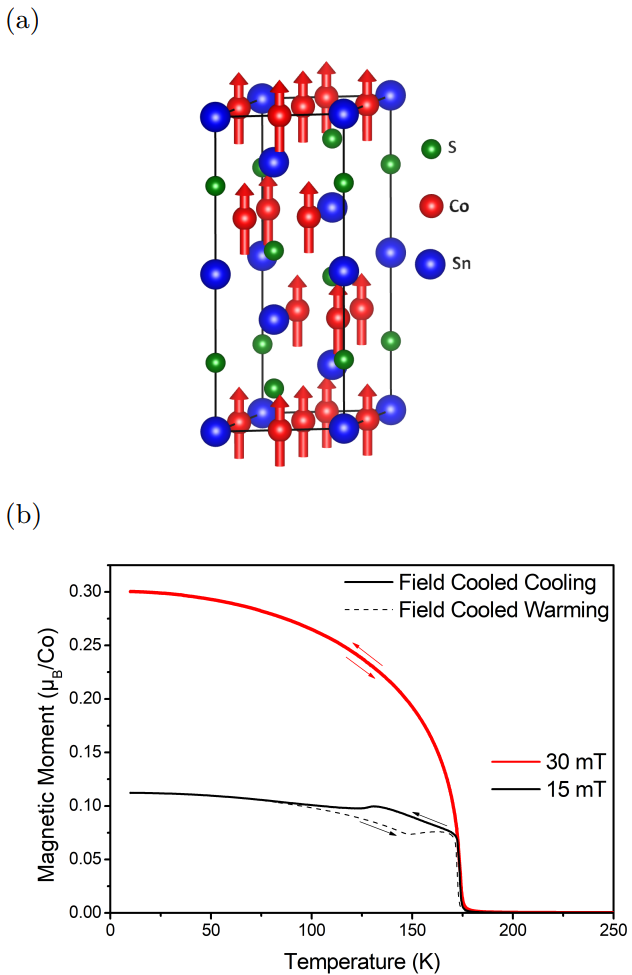}
    \caption{\label{fig:introduction}\justifying{\small{(a) Crystal structure of Co$_3$Sn$_2$S$_2$  with arrows showing the orientation of spins in the ordered phase. (b) Magnetization as a function of temperature at several magnetic fields. Above 22 mT, magnetization in the ferromagnetic phase is featureless. But when the sample is cooled down in presence of a magnetic field smaller than this threshold field, there is an anomaly and a hysteresis, which extends down to $\simeq 110$ K. }}}
\end{figure}

First synthesized as a ternary chalcogenide with Shandite structure \cite{ZabelWandingerRange+1979+238+241}, Co$_{3}$Sn$_{2}$S$_{2}$ became a subject of tremendous attention after its identification as a ferromagnetic Weyl semi-metal \cite{liu2018giant,wang2018large}. It crystallizes in a rhombohedral structure with R$\Bar{3}$m space group (n°166). The cobalt atoms form kagome sheets in the ab plane, which are separated by blocks of Sn and S atoms (see Figure \ref{fig:introduction}a). It magnetically orders below $T_C$ $\approx 175$ K with a saturation moment of $\approx 0.3$ $\mu_B$ per Co atom \cite{VAQUEIRO2009513}, and with the easy axis residing along the c-axis. \textit{Ab initio} band calculations \cite{Dedkov_2008}, as well as photoemission \cite{Holder2009} and M\"ossbauer experiments \cite{Schnelle2013} identified it as a ferromagnetic half-metal. It is also a semi-metal with an equally low concentration of electrons and holes ($n$ = $p$ $\simeq 8.7 \times 10^{-19}$cm$^{-3}$ \cite{ding2019intrinsic}). Thanks to such a low carrier density (comparable to elemental antimony, where $n$ = $p$ $\simeq 6.6 \times 10^{-19}$cm$^{-3}$), mobility is sufficiently large to detect quantum oscillations and experimentally confirm  the theoretically computed  Fermi surface, consisting of two electron-like and two hole-like and multiply degenerate sheets \cite{Ding_2021}. 

The low carrier density implies that each mobile electron is shared by several hundred formula units of Co$_{3}$Sn$_{2}$S$_{2}$. This distinct feature leads to an exceptionally large anomalous Hall angle \cite{liu2018giant}. Indeed, although the anomalous Hall conductivity of  Co$_{3}$Sn$_{2}$S$_{2}$ ($\sigma^A_{xy}$(2K) $\simeq1200 (\Omega$cm$)^{-1}$ \cite{liu2018giant,ding2019intrinsic}) falls below what is seen in CoMn$_2$Ga ($\sigma^A_{xy}$(2K) $\simeq2000(\Omega$cm$)^{-1}$ \cite{Sakai2018}), the anomalous Hall angle attains a record magnitude ($\frac{\sigma^A_{xy}}{\sigma_{xx}}$(120 K) $\simeq 0.2$) in Co$_{3}$Sn$_{2}$S$_{2}$ \cite{liu2018giant}. Another consequence of high mobility is seen in the Nernst response. In contrast with  low-mobility topological magnets (like Mn$_3$X (X=Sn, Ge) in which the Nernst effect is purely anomalous \cite{Li2017,Ikhlas2017,xu2020-2}), Co$_{3}$Sn$_{2}$S$_{2}$ has a sizeable ordinary Nernst response in addition to the anomalous component. Their ratio can be tuned by changing the concentration of impurities \cite{ding2019intrinsic}.

\begin{figure*}
    \centering
    \includegraphics[width=0.93\linewidth]{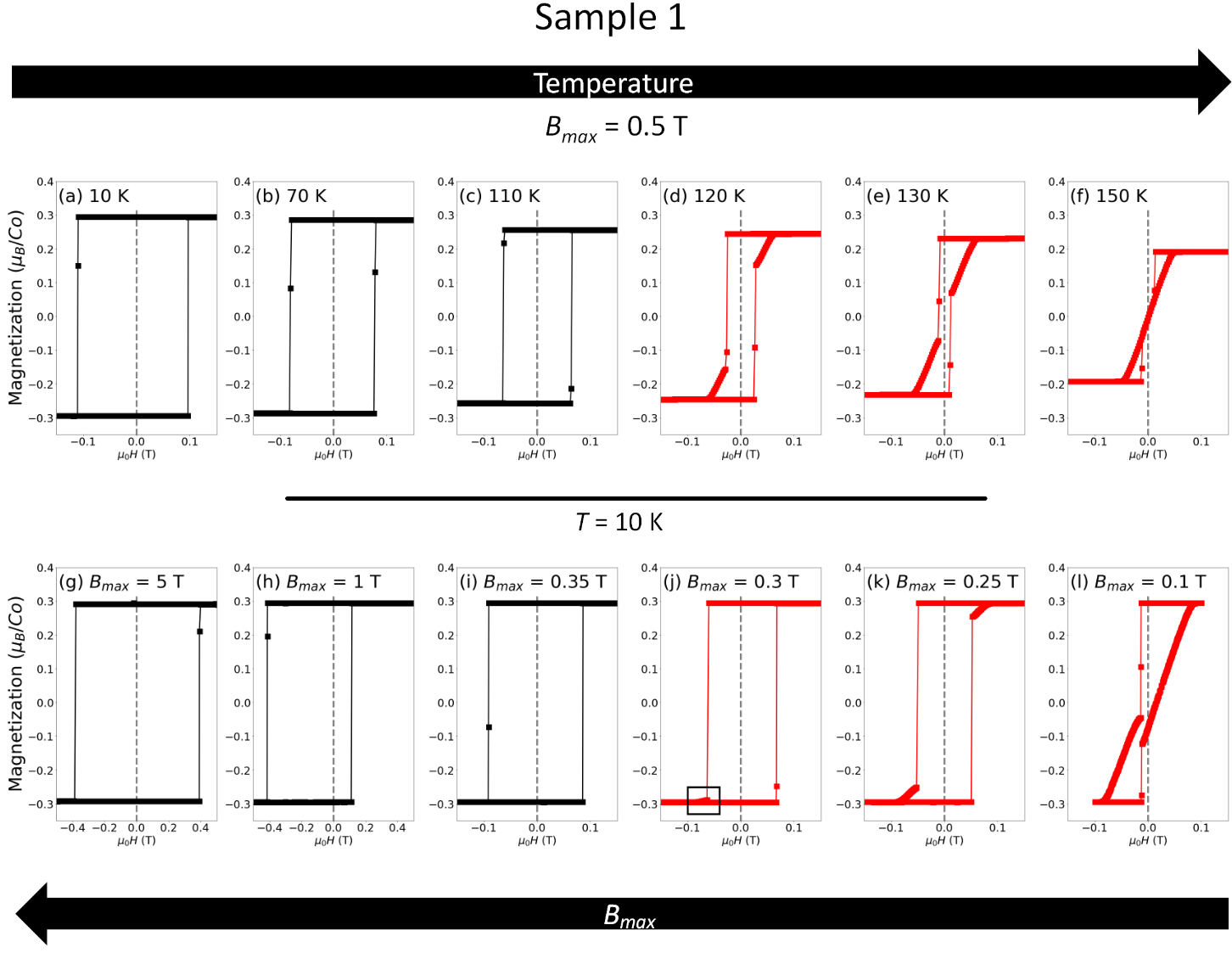}
    \caption{\label{fig:loop}\justifying{\small{\textbf{Shape of the magnetization hysteresis loops} (a)-(f) Hysteresis loops at 10 K, 70 K,110 K, 120 K, 130 K, 150 K, all with an identical maximum sweeping magnetic field ($B_{max}$ = 0.5 T). Note the emergence of a bow-tie feature above 120 K.  (g)-(l) Hysteresis loops at T = 10 K with different maximums sweeping fields ($B_{max}$ = 5 T, 1 T, 0.35 T, 0.3 T, 0.25 T and 0.1 T). When $B_{max}$ becomes smaller than $0.35$ T, a bow-tie feature emerges.}}}
\end{figure*}

Together with Fe$_3$Sn$_2$ \cite{Ren2022}, Co$_{3}$Sn$_{2}$S$_{2}$ belongs to the restricted family of kagome ferromagnets. However, several recent experimental studies  \cite{guguchia2020tunable,zhang2021unusual,lee2022observation,soh2022magnetic,neubauer2022spin,kassem2017low,lachman2020exchange,vzivkovic2022unraveling,avakyants2023evidence,noah2022tunable,shen2022anomalous,zhang2022hidden} suggested that the magnetic ordering in Co$_{3}$Sn$_{2}$S$_{2}$ is not purely ferromagnetic. In addition to the Curie temperature ($T_C\simeq 175$ K), there is an additional temperature scale, $T_A$.  Muon spin-rotation ($\mu_{SR}$) measurements \cite{guguchia2020tunable} suggested the presence of an in-plane anti-ferromagnetic component emerging above 90 K that occupies an increasing volume fraction with warming and becomes dominant above 150 K. A Kerr microscopy study \cite{lee2022observation} reported that near $T_A\simeq$ 135 K, domain wall mobility goes through a deep minimum, pointing to a phase transition within the domain walls. A recent neutron scattering study \cite{soh2022magnetic} found no evidence for antiferromagnetism in the magnetically ordered state and attributed the features observed near 125 K to a reduction of ferromagnetic domain size. Another study \cite{neubauer2022spin} found that an anti-ferromagnetic component emerges with indium doping in Co$_{3}$Sn$_{2}$S$_{2}$. Lachman \textit{et al.} \cite{lachman2020exchange} found that the hysteresis loop of the anomalous Hall effect is not centered around zero field, a feature reminiscent of the so-called ``exchange bias'' effect in ferromagnet/antiferromagnet bilayers \cite{Stamps_2000}. Moreover, they found that the magnetization hysteresis loop, which has a rectangular shape at low temperature, displays a ``bow-tie'' structure above $T_A\simeq$ 125 K. This led them to suggest the existence of a spin-glass phase. Zivkovic \textit{et al.} \cite{vzivkovic2022unraveling} reported a similar change in the shape of the magnetic hysteresis loop and diagnosed a phase transition at $T_A\simeq$ 128 K associated with a change in the canting angle of the magnetic moment away from the c-axis. On the other hand, Avakyants \textit{et al.} \cite{avakyants2023evidence}, employing a  First Order Reversal Curves (FORC) analysis, concluded that two independent magnetic phases coexist below T$_C$. Noah \textit{et al.} \cite{noah2022tunable} reported that the exchange bias found in this system \cite{lachman2020exchange} can be tuned by changing the prior history of the sample. 

Here, we present a systematic study of magnetization as a function of temperature, magnetic field and prior magnetic history in Co$_3$Sn$_2$S$_2$ single crystals and identify the origin of $T_A$.  We find that the ``bow-tie'' shape of the hysteresis loop \cite{lachman2020exchange,shen2022anomalous,pate2023field} is not restricted to temperatures exceeding $T_A$. Even at low temperature, when the maximum field of opposite polarity visited by the sample ($B_{max}$) is sufficiently small, the hysteresis has a bow-tie shape \cite{noah2022tunable}. Not only the shape of the loop but also other features such as the threshold field for flipping spins ($B_{0}$) and the asymmetry between opposite field polarities ($B_{0+}\neq B_{0-}$), the exchange bias, depend on  $B_{max}$. Thus, the system  has a memory of the previously visited $B_{max}$.  We identify  a distinct small spin population as the driver of this memory. They keep their polarity even when the magnetic field is almost an order of magnitude larger than the coercive field for most ($>99\%$) of the spin population. A detailed study of this memory effect leads us to conclude that T$_A$ is not a thermodynamic phase transition, but a crossing point between boundaries in the (temperature, field) plane. One boundary separates memory-less and memory-full regimes. The other frontier determines the shape of the hysteresis loop and multiplicity of domains. When a single-domain state is abruptly replaced by a single-domain state of opposite polarity, the loop has a rectangular shape. A bow-tie shape emerges when  the reversal has a multi-domain interlude. The existence of more than one type of ordered spins  may  be due to bulk/surface dichotomy \cite{avakyants2023evidence}. Finally, by performing local magnetometry studied with miniature Hall probes, we show that when magnetization is not uniform, features associated with thermodynamics of small systems may emerge at length scales as large as a few microns. 

\section{Results}
Figure \ref{fig:introduction}b shows the temperature dependence of magnetization in one of our samples. Magnetization is enhanced below the Curie temperature of 175 K and saturates to 0.3 Bohr magneton, $\mu_B$, per Co atom  at low temperatures. Inside the ferromagnetic state, an anomaly and a hysteresis in temperature are detectable, which both disappear when the applied magnetic field becomes large enough to saturate magnetization.

\subsection{Magnetization Hysteresis loops}

Figure \ref{fig:loop} a-f shows the evolution of the magnetization hysteresis loop as a function of temperature in a Co$_{3}$Sn$_{2}$S$_{2}$ single crystal. These loops were obtained by sweeping the field between -0.5 T and +0.5 T, corresponding to $B_{max}$ = 0.5 T. At low temperatures (Fig \ref{fig:loop} a,b,c), the hysteresis loop looks like a rectangle as in a hard magnet. The magnetic field suddenly flips all spins at well-defined thresholds identified as $B_{0+}$ and $B_{0-}$. Let us define $B_{0}=(B_{0+}-B_{0-})/2$, the average spin-flip field. At T $> 120$ K (Fig \ref{fig:loop} d,e,f,)  the hysteresis loop is no more rectangular. This implies that all spins do not flip at $B= B_{0\pm}$. The jump in magnetization is followed by a smooth and almost field-linear variation of magnetization. This is the 'bow-tie' hysteresis shape \cite{lachman2020exchange}. 

The evolution seen in  Figure \ref{fig:loop} a-f is similar to what was reported by Lachman \textit{et al.} \cite{lachman2020exchange}, who found that the hysteresis loop acquires a `bow-tie' shape  above a threshold temperature. The only difference is that our threshold temperature ($T_A= 115$ K $\pm$ 5 K) is lower than theirs ($T_G=125$ K). This difference will be explained at the end of this paper. Another feature which appears at low temperature is a genuine asymmetry between positive and negative orientations : $B_{0+}\neq B_{0-}$ \cite{lachman2020exchange}, which is particularly visible in Figure \ref{fig:loop}h. 

Panels g-l in figure \ref{fig:loop} show the hysteresis loops at 10 K with different maximum sweeping fields, $B_{max}$. The evolution is similar to the one induced by warming. When $B_{max}=5$ T, the magnetization loop is rectangular. Decreasing $B_{max}$ reduces $B_{0}$, in agreement with what was previously reported \cite{noah2022tunable}. For sufficiently small $B_{max}$ (that is, when $B_{max} <$ 0.35 T), the hysteresis loop acquires a bow-tie shape. The emergence of bow-tie shape and low values of $B_{0}$ are concomitant. We refer to the amplitude of $B_{0}$ below which the hysteresis loop displays a bow-tie feature as $B_0^{bt}$. 
 
Thus, at low temperature, the shape of the hysteresis loop and the amplitude of $B_0$ both depend on $B_{max}$. In other words, the amplitude of magnetization at a given magnetic field does not exclusively depend on temperature and magnetic field, but also on the magnetic field applied in the past. If the latter is not large enough, a memory persists. Memory formation in condensed matter is defined as an `ability to encode, access, and erase signatures of past history in the state of a system' \cite{Keim2019}. The present case is reminiscent of another topological magnet, namely Mn$_3$X (either with X=Sn \cite{li2019chiral} or X=Ge \cite{Xu2020}). However, as we will see below, here the information is stocked not in the domain walls between antiferromagnetic domains, but in a secondary spin population.  

\begin{figure*}
    \centering
    \includegraphics[width=0.8\linewidth]{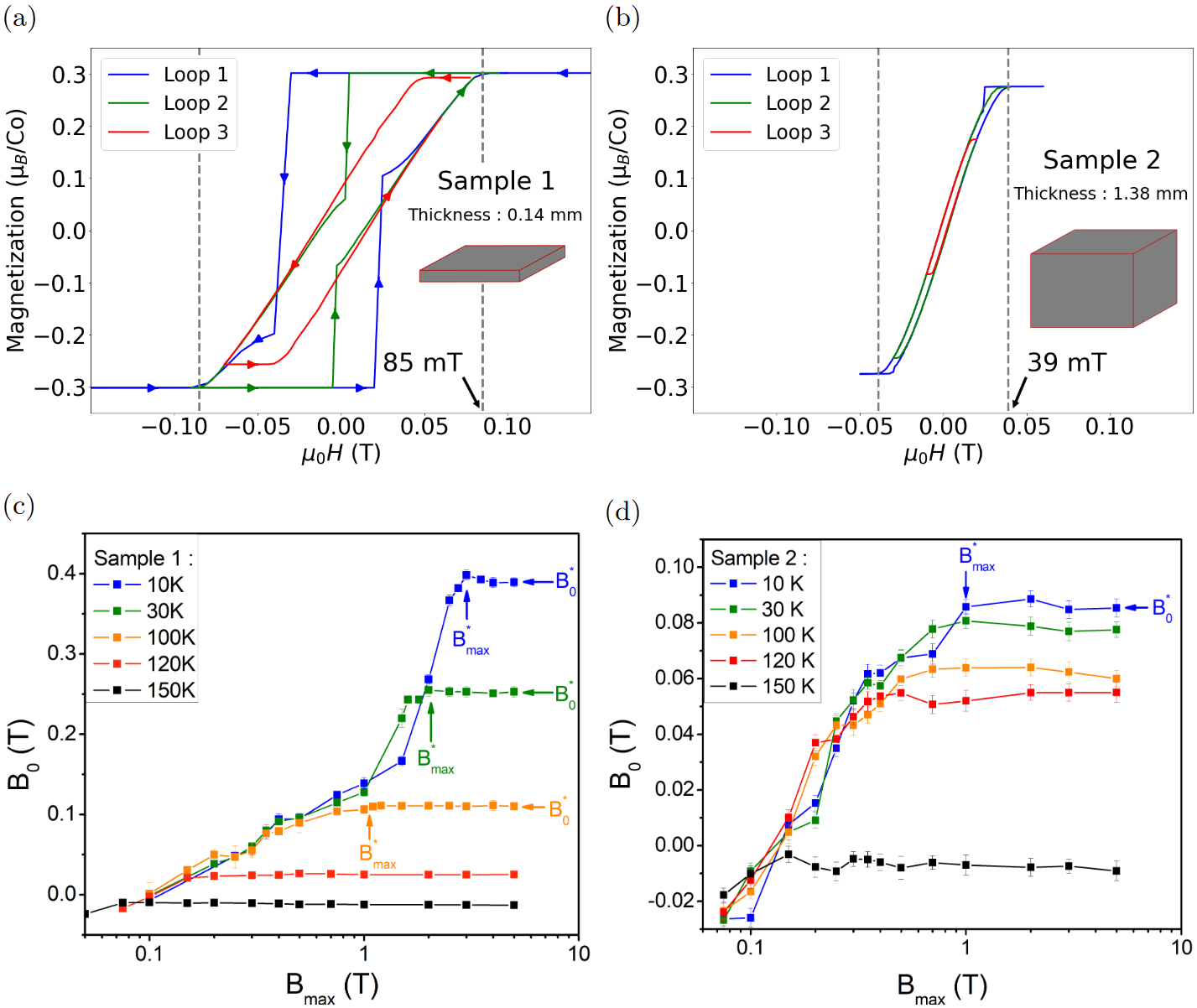}
     \caption{\justifying{\small{(a) Three Magnetization loops at 10 K. Loop 1 (blue): starting field at 0.2 T. Loop 2 (green): starting field at 0.1 T. Loop 3 (red): starting field at 0.08 T. (b) Loops for a thicker sample. Loop 1 (blue): starting field at 0.06 T. Loop 2 (green): starting field at 0.04 T. Loop 3 (red): starting field at 0.02 T. Bow-tie features tend to two parallel lines. (c) $B_0$ as function of $B_{max}$ at different temperatures for sample 1; (d) Same for sample 2. At each temperature, $B_0$ initially increases linearly with increasing $B_{max}$. It eventually saturates to a constant value. This threshold sweeping field, $B_{max}^\star$, is shown by arrows. The saturated amplitude of $B_0$, called $B_0^\star$ is also shown.  Both $B_0^\star$ and $B_{max}^\star$ decrease with increasing temperature.}}}
    \label{fig:tail}
\end{figure*}

\subsection{Origin of the Bow-tie shape}
Multiplicity of magnetic domains in Co$_3$Sn$_2$S$_2$, which occurs when the amplitude of magnetization is below its peak value of $M_s\simeq 0.3$ $\mu_B$/Co, has been probed by microscopic techniques \cite{lee2022observation,sugawara2019magnetic,howlader2020domain}. To identify the origin of $B_0^{bt}$ and the change in the shape of the hysteresis loop across this threshold, we scrutinized  hysteresis loops with a very small ($<0.2$ T) $B_{max}$, leading to a magnetization lower than $M_s$. 

Figure \ref{fig:tail}a illustrates the  variation of magnetization with applied magnetic field during three successive hysteresis loops where the amplitude of $B_{max}$ is incrementally reduced after each loop. The first two loops (blue and green) have a bow-tie shape: magnetization is first flat, then abruptly drops (or jumps) and then shows a steady drift towards its saturated value with a slope tending to be independent of $B_{max}$. During this steady drift, the system hosts multiple magnetic domains. In the third loop (red), $B_{max}$ is so low that abrupt jumps (or drops) vanish. Note that the slope of magnetization in the red loop is similar to the slope of magnetization in the green and blue loops which presents a bow-tie shape. This slope, which does not depend on $B_{max}$, sets $B_0^{bt}$. Dividing $\mu_0$M$_s$ by  $dM/dH$ yields 85 mT (Figure \ref{fig:tail}a), close to the threshold $B_0^{bt}$ revealed in the transition between panels i and j of Figure \ref{fig:loop}. 

Magnetization loops for a thicker sample (Figure \ref{fig:tail}b) are similar, but there is a quantitative difference. The magnetization slope is steeper in the thicker sample, which has almost a cubic shape. $M_s$ is identical in the two samples and therefore the threshold field is reduced to 39 mT in this thicker sample. Thus, with increasing thickness, the multi-domain window becomes narrower, the magnetization slope becomes steeper and $B_0^{bt}$ is reduced. 

The difference in the demagnetizing factors of the two samples provides a quantitative account of this thickness dependence. 
As seen in table \ref{tab:D}, $D$, the demagnetizing factor calculated by using the formula for a rectangular prism \cite{aharoni1998demagnetizing}, is very close to $\frac{dH}{dM}$, the inverse of the magnetization slope measured in the experiment.

\begin{table*}[ht]
    \centering
    \begin{tabular}{|P{4em}|P{9em}|P{6em}|P{5em}|P{5em}|}
        \hline
         & Dimensions ($mm^3$) & Aspect ratio & $D$ & $dH/dM$ \\
         \hline
        Sample 1 & 1.21 $\times$ 0.89 $\times$ 0.14 & 0.13 & 0.76 & 0.69 \\
        \hline
        Sample 2 & 1.93 $\times$ 1.20 $\times$ 1.38 & 0.88 & 0.35 & 0.38 \\
        \hline
    \end{tabular}
    \caption{\justifying{Samples dimensions and aspect ratio. For both samples, the calculated demagnetizing factor,$D$ , is close to the measured $dH/dM$ (the inverse of the magnetization slope) when $B<B_0^{bt}$ implying $B_0^{bt}=\mu_0D$M$_s$. }}
    \label{tab:D}
\end{table*}

When the magnetic field becomes equal to $B_{0\pm}$, an energy barrier is crossed and spins flip to the opposite orientation. If $|B_{0\pm}|\geqslant\mu_{0}DM_s$, the spin-flip is total and the loop is rectangular. On the other hand, if  $|B_{0\pm}|\leqslant\mu_{0}DM_s$, spin-flip is partial and the loop has a bow-tie shape, because a multi-domain configuration is stable thanks to the demagnetization energy. This leads us to $B_{0}^{bt}=\mu_{0}DM_s$, in agreement with the experimental observation.

\subsection{Temperature dependence of the memory effect}
We saw that the magnitude of $B_0$ depends on the maximum sweeping field, $B_{max}$. Figure \ref{fig:tail}c illustrates this dependence for different temperatures in a semi-log plot. At each temperature, the initial increase $B_0$, which is roughly linear in $B_{max}$ ends by saturation to a constant value, which we dub $B_{0}^\star$. Let us call B$_{max}^\star$ the magnitude of $B_{max}$ above which $B_0=B_{0}^\star$. As one can see in the figure, both $B_{0}^\star$ and $B_{max}^\star$ steadily decrease  with increasing temperature. The picture drawn by this data is following: when $B_{max}< B_{max}^\star$, the system has a memory, which shows itself in the magnitude of $B_0$. Spin flip occurs at a threshold magnetic field, which depends on the previously visited field. When $B_{max}>B_{max}^\star$, there is no such memory and $B_0 (T)=B_{0}^\star(T)$ is independent of previous history.

Figure \ref{fig:tail}d presents the same data for the thicker sample ($\#2$). The behavior is qualitatively similar: After an initial increase, $B_0$ saturates at a temperature dependent magnitude. Note, however, that the absolute value of $B_0^\star$ is much smaller in the thicker sample. At low temperature, $B_0^\star\simeq 0.4$ T in sample 1 and $B_0^\star\simeq 0.09$ T in sample 2. It is noteworthy that, for both samples, $B_{max}^\star/B_{0}^\star \approx 8$ and this ratio does not show any strong temperature dependence. This indicates that the temperature-induced decrease in both field scales is similar.

 $B_0^\star$ is the coercive field of the system when the memory is erased. As expected \cite{garcia1998influence}, it decreases with increasing  temperature. Its amplitude in the zero-temperature limit is much smaller than the magneto-crystalline anisotropy field, i.e. to the in-plane magnetic field needed to saturate magnetization. The latter is as large as $\sim$ 23 T \cite{shen2019anisotropies}. This difference makes Co$_{3}$Sn$_{2}$S$_{2}$ another example of what is known as 'Brown's coercivity paradox' \cite{hartmann1987origin}.
Experiments have found that the coercivity is often much smaller than the lowest bound  expected according to the magneto-crystalline anisotropy \cite{coey2010magnetism}. It has been shown that large demagnetizing fields developed near sharp corners play a significant role in setting coercivity \cite{hartmann1987origin} and imperfections can reduce the expected coercive field \cite{aharoni1960reduction}.

\begin{figure}
    \centering
    \includegraphics[width=0.95\linewidth]{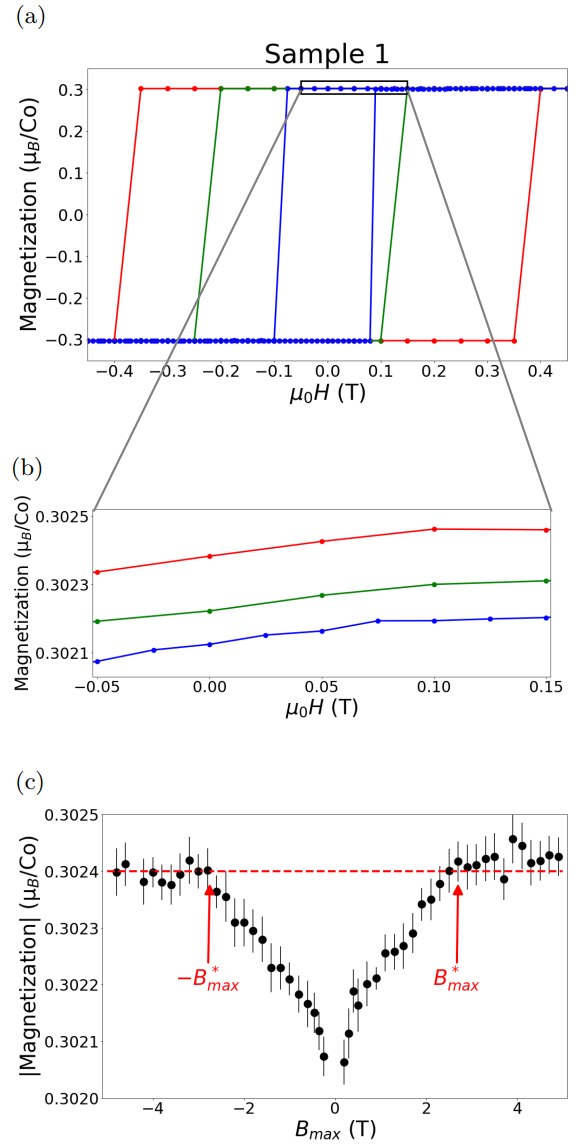}
    \caption{\justifying{\small{Memory effect in "saturated" magnetization, at 10 K. (a) Three magnetization loops, red : $B_{max}$ = 4.8 T, green: $B_{max}$ = 1.5 T and blue: $B_{max}$ = 0.5 T. (b) Zoom in positive magnetization at low field.  (c) $B_{max}$ dependence of the absolute value of magnetization  at $\pm$ 50 mT (depending on $B_{max}$ sign).}}}
    \label{fig:Ms_B_max_total}
\end{figure}

Thus, when the maximum sweeping field $B_{max}$ becomes lower than a  threshold ($B_{max}^\star$), $B_0$ becomes lower than its peak value $B_0^\star\simeq 0.4$ T. Moreover, this is also the case of the difference between $|B_{0+}|$ and $|B_{0-}|$, which becomes significantly larger than our experimental margin. Thus, the memory effect tunes the exchange bias, too (See the Supplementary Materials for details \cite{SM}).

\subsection{Zooming on saturated magnetization}
Given the content of this memory, one may suspect that when it is not erased, the energy barrier between two single-domain states is attenuated. This may be caused by the  presence of a secondary spin component or population which modifies the overall energy landscape and attenuates the height of the barrier. A careful examination of saturated magnetization at the end of a hysteresis loop confirms this. 

Figure \ref{fig:Ms_B_max_total}a displays three loops all at 10 K with three different endings ($B_{max}$=0.5 T; 1.5 T and 4.8 T). The figure shows that $B_0$ increases with increasing $B_{max}$, as we saw above. At first sight, magnetization appears to saturate at the same amplitude. However, this is not the case. Figure \ref{fig:Ms_B_max_total}b is a zoom on the three curves near the maximum magnetization. One can see that there is a small, yet finite difference between the three curves. With increasing $B_{max}$, the amplitude of magnetization at the end of a `rectangular' loop is larger. We carefully documented the dependence of spontaneous magnetization at the end of a loop (measured at $B=\pm$ 0.05 T) on the amplitude of the sweeping magnetic  field. 

Figure \ref{fig:Ms_B_max_total}c shows the result. The spontaneous magnetization at the end of a loop increases with increasing $|B_{max}|$ before saturating to a constant value when $B_{max}$ becomes equal to $B_{max}^\star$. The detected increase of magnetization between $B_{0}^\star$ (the end of the loop) and its eventual saturation above $B_{max}^\star$ is tiny ($\approx$ 0.1\%), but larger than our experimental margin. This observation has an important implication: when the sample has not visited a sufficiently large $B_{max}$, it hosts a small population of spins whose magnetization does not correspond to the polarity of majority spins. This population is where the memory is stored. The existence of a $B_{max}^\star$ (roughly 8 times larger than B$_0^\star$) is caused by this secondary spin population whose coercive field is larger and much more broadly distributed than the coercive field of the majority spins. The secondary spins, presumably three orders of magnitude more dilute than the principal population, may be situated either at the surface of the sample or at off-stoichiometric sites.

\begin{table*}
    \centering
    \begin{tabular}{|c|c|c|c|}
        \hline
          Field scale & Amplitude  (T)  & Definition  \\
         \hline
        $B_{0}^\star$ (T) & 0.4 & Spin-flip field in absence of memory (coercivity  of main spin population) \\
         \hline
        $B_{max}^\star$ (T) & 3 & Sweeping field above which no memory persists (maximum coercivity  of secondary spin population) \\
          \hline
        $B_{0}^{bt}$ (T) & 0.085 & Hysteresis becomes bow-tie when the spin-flip field falls below this threshold \\
         \hline
        $B_{max}^{bt}$ (T) & 0.35 & Hysteresis becomes bow-tie when the magnetic field is swept below this threshold\\
         \hline
    \end{tabular}
    \caption{\justifying{Four distinct field scales identified in this study. The amplitudes are given for sample 1 at $T=10$ K. $B_{0}^\star$ and $B_{0}^{bt}$ refer to spin-flip fields, at which the magnetic field shows an abrupt jump. $B_{max}^{bt}$ and $B_{max}^\star$  refer to maximum sweeping field tuning the spin-flip field, $B_0$. $B_{0}^{bt}$ and $B_{max}^{bt}$ depend on the demagnetizing factor and show little dependence on temperature. $B_{max}^\star$  and  $B_{0}^\star$ both decrease with warming. }}
    \label{tab:F}
\end{table*}
\begin{figure}
    \centering
    \includegraphics[width=0.98\linewidth]{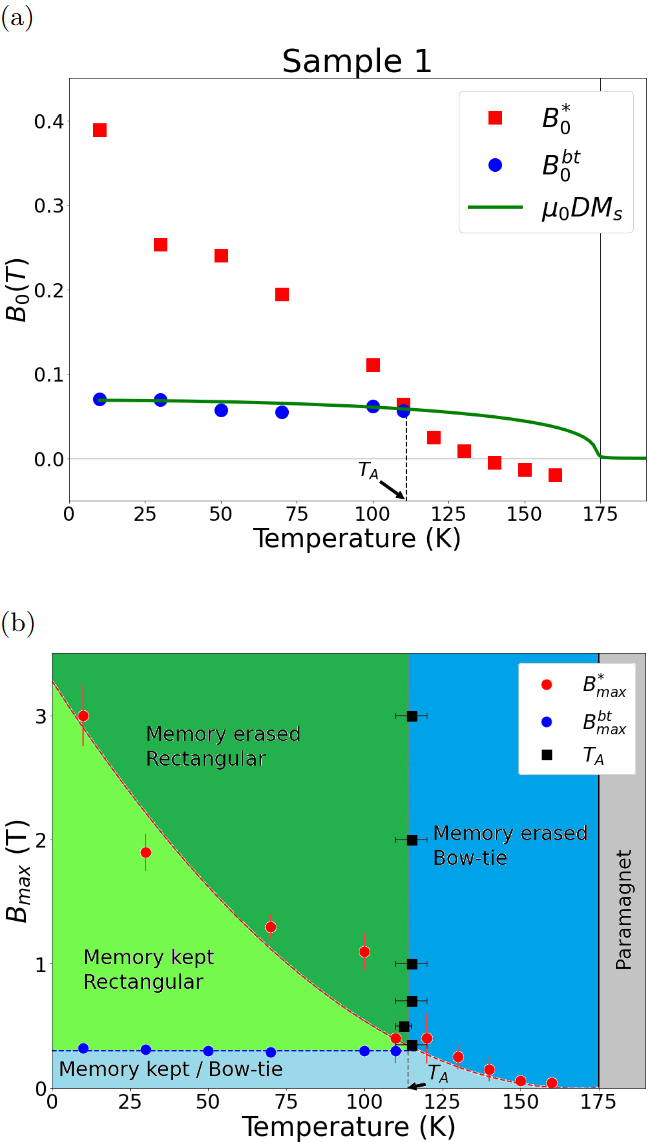}
    \caption{\label{fig:diag}\justifying{\small{(a) $B_{0}^\star$ and $B_{0}^{bt}$ versus temperature. The measured $B_{0}^{bt}$ tracks $\mu_0DM_s$ (with $D=0.76$), which is represented by the green solid line.  $B_{0}^\star$ and $B_{0}^{bt}$ cross each other at $T_A$. Above this , hysteresis loops can only have a bow-tie shape for whatever $B_{max}$, because of the B$_{0}^\star<B_{0}^{bt}$ inequality. (b) B$_{max}^\star$ and $B_{max}^{bt}$ versus temperature. Dashed lines are guides to the eye. $T_A=$ 115 $\pm$ 5 K, corresponds to the crossing point of B$_{max}^\star$ and $B_{max}^{bt}$.}}}
\end{figure}
\begin{figure*}
    \centering
    \includegraphics[width=0.93\linewidth]{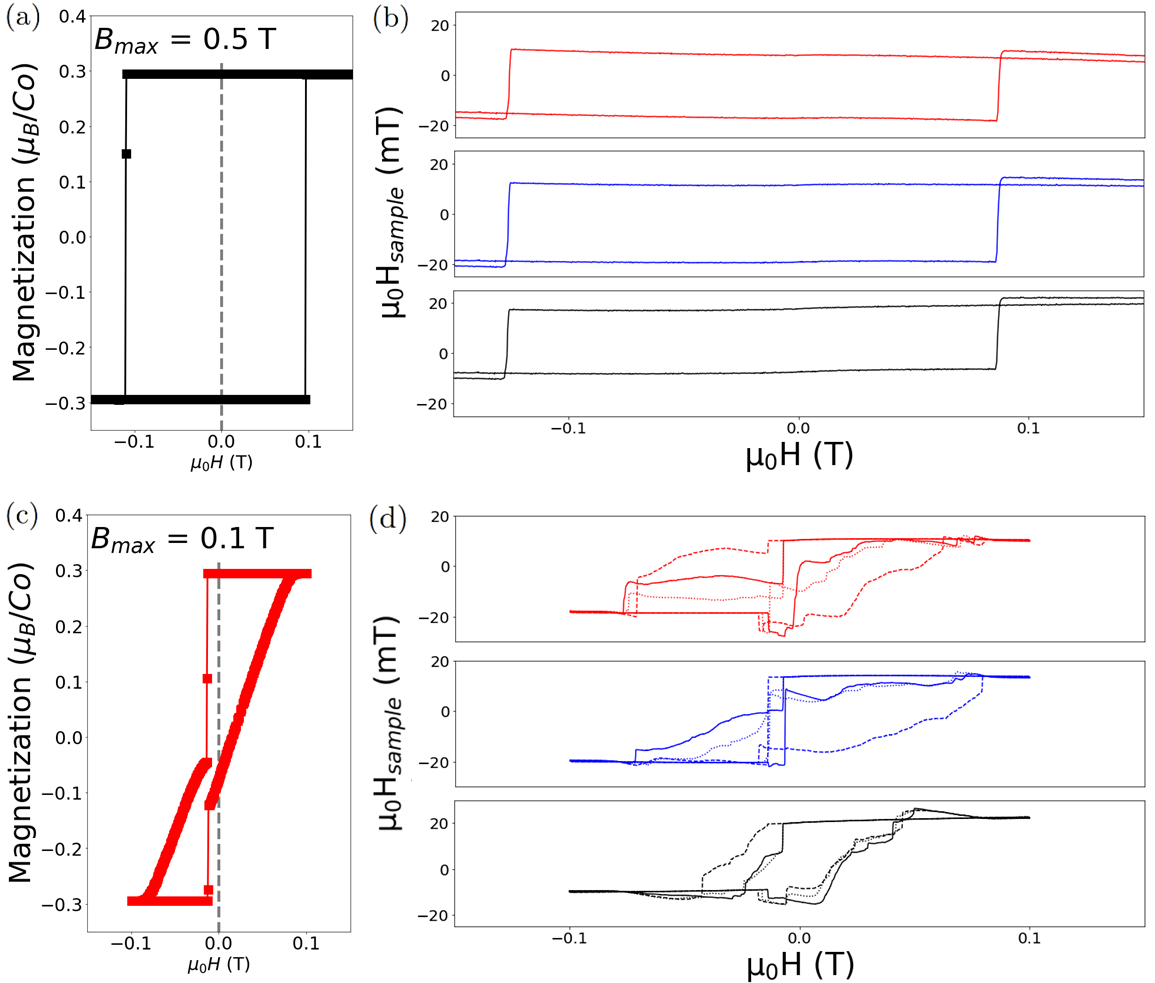}
    \caption{\label{fig:sonde_hall}\justifying{\small{Local and global magnetization. (a) Magnetization of the whole sample, measured with a vibrating-sample magnetometer, at 10 K for $B_{max}$ = 0.5 T. (b) Local magnetization, measured in identical conditions by an array of 2DEG micron-sizes Hall sensors put on the surface  of the sample. The three curves represent the local magnetic field at three different positions of the sample. (c) Magnetization of the whole sample at 10 K for $B_{max}$ = 0.1 T. Hysteresis displays a bow-tie shape.  (d) Local magnetization measured in similar conditions. Each color represents a position on the sample (three consecutive hysteresis loops were measured.)}}}
\end{figure*}
\section{Discussion}
Having identified four different field scales (See table \ref{tab:F}), let us now turn our attention to the phase diagram.

\subsection{Origin of T$_A$, the additional temperature scale}

Figure \ref{fig:diag}a shows the evolution of $B_{0}^\star$ and $B_0^{bt}$ with temperature. $B_0^{bt}$, the threshold field  for bow-tie shape, is almost flat and its absolute value coincides with $\mu_0DM_s$. In contrast, $B_{0}^\star$, the saturated $B_0$, is temperature dependent and rapidly decreases with increasing temperature.  $T_A$ is the temperature at which $B_{0}^\star$ and $B_0^{bt}$ cross each other. When $B_{0}^\star$ falls below $B_0^{bt}$, whatever the sweeping field $B_{max}$, one finds $B_{0} <B_0^{bt}$. This makes a bow-tie shape unavoidable. Thus, no thermodynamic phase transition occurs at $T_A$. This temperature threshold arises as a result of the crossing between two boundaries.

This is further illustrated in Figure \ref{fig:diag}b, a representation of the evolution of $B_{max}^\star$  and $B_{max}^{bt}$ in the (field, temperature) plane. When $B_{max} > B_{max}^\star$, the system has no memory (that is, the shape of hysteresis loop does not depend on the past history) and when $B_{max} < B_{max}^\star$, there is a memory. $B_{max}^\star (T)$ is the boundary between two states with and without memory. A second boundary is defined by $B_{max}^{bt}$. When $B_{max}$ is larger than this threshold, the hysteresis loop is rectangular. Since $B_{max}/B_0$ ratio does not change with temperature,  $T_A$, the temperature at which $B_{max}^{bt}= B_{max}^\star$ is similar to what can be seen in Figure \ref{fig:diag}a. Since $B_0^{bt}$ and $B_{max}^{bt}$ depend on the demagnetizing factor and the sample morphology, the horizontal lines in these figures are expected to vary from sample to sample with the change in the aspect ratio. This would explain the fact that the reported secondary temperature differs from one study to another.

Thus, the magnetic order in Co$_{3}$Sn$_{2}$S$_{2}$ remains ferromagnetic with spins oriented along the c-axis. However, in presence of a magnetic field oriented along the c-axis, as a function of temperature and magnetic field (both at present and in the past), multiple meta-stable configurations can arise. What distinguishes them is the polarity in different sub-sets of spin population, despite the global ferromagnetic order. 

\subsection{Thermodynamic limit for well-defined magnetization}
One manifestation of this meta-stability is the contrast between local and global magnetization, shown in Figure \ref{fig:sonde_hall}. When the system is in the rectangular regime, local magnetization, measured with micron-sized Hall sensors (see supplementary material for details \cite{SM}), is very similar to the magnetization of the whole sample. In other words, the passage between single-domain regimes of opposite polarities is almost identical everywhere in the sample (Figure \ref{fig:sonde_hall} a \& b). On the other hand, when the sample is multi-domain, while global magnetization presents a smooth and reproducible slope (Figure \ref{fig:sonde_hall}c), local magnetization is not reproducible from one sweep to another (Figure \ref{fig:sonde_hall}d). This confirms that when the memory is not erased, the energy landscape is not smooth \cite{Wales2001}. There are numerous competing spin configurations, spatially distinct over a micrometer, but with similar global magnetization. Our case emerges as a platform for studying thermodynamics of small systems \cite{hill1994thermodynamics}.

\subsection{Possibly related phenomena in other magnetic solids}
Let us note that Co$_3$Sn$_2$S$_2$ is not the first case of multiple spin populations. In SrRuO$_3$ thin films \cite{Ding2023,Kar2023} with a thickness of few unit cells, there is an additional peak in the Hall response. It has been proposed that it is caused by contributions of opposite
signs from two distinct magnetic regions with different saturation magnetizations \cite{kimbell2020two}. The two contributions have different coercive fields, but this difference is far below the order of magnitude difference we see in our case. A similar observation has been reported in NiCo$_2$O$_3$ thin films \cite{zhou2023observation}.

These observations indicate that the existence of distinct spin populations with different coercive fields in a magnet may be more common than previously thought.

\section{Conclusion}
In summary, we investigated the evolution of the magnetization hysteresis loop in Co$_{3}$Sn$_{2}$S$_{2}$  with temperature and with the maximum swept magnetic field,  $B_{max}$. We found that, at each temperature, increasing $B_{max}$ leads to an enhancement of the coercive field up to a saturation value, $B^\star_{max}$. In addition, the amplitude of the saturated magnetization displays a small, yet significant, dependence on $B_{max}$. This suggests the presence of a small secondary spin population with a coercive field larger than that of the main population. The memory of the last  $B_{max}$ is stored by this minority spins, which do not flip if the sweeping field is lower than  $B^\star_{max}$.

A temperature scale, $T_A$,  distinct from the Curie temperature, was identified by several previous studies. It was suggested that it corresponds to a thermodynamic phase transition within the magnetically ordered state. This secondary phase was suggested to be in-plane antiferromagnetism \cite{guguchia2020tunable}, spin glass \cite{lachman2020exchange} or an anomaly in domain wall mobility \cite{lee2022observation}. According to our study, T$_A$ does not correspond to a thermodynamic phase transition, but to a crossing point between meta-stable states. The two boundaries which cross at T$_A$ separate  regimes with and without memory and regimes which are single-domain and multiple-domain. The origin of the two distinct coercive fields corresponding to two distinct spin populations emerges as a puzzle to be addressed by future studies.

\section{Acknowledgements}
This study was supported  by Jeunes Equipes de l'Institut de Physique du Coll\`ege de France, and by a grant attributed by the Ile de France regional council. C. M. acknowledges a PhD scholarship granted by CNRS.

\section{Methods}
Crystals of Co$_{3}$Sn$_{2}$S$_{2}$ were grown by self-flux method as detailed previously \cite{Ding_2021}.

Magnetization was measured using a Quantum Design MPMS in Vibrating Sample Magnetometer (VSM) mode with magnetic field applied along the crystalline c-axis with a quartz sample holder.

The hysteresis loops displayed in figures 2 and 3 c-d, were obtained according to the following protocol :
\begin{itemize}
    \item Set the temperature.
    \item Reduce the remanent field by sweeping the applied field with gradually decreasing $|B_{max}|$ until finding that magnetization is not saturated anymore. For instance, the field was swept first from -0.5 T, to 0.2 T then to -0.1 T, then to 0.05 T, and finally to 0 T.
    \item Loops were measured consecutively, starting from the smallest to the largest $B_{max}$ without additional delay between steps of measurement.
    \item Loops measurement were done by initially setting the external field to $B_{max}$ (with a sweep rate of 100 Oe/s). Then the external field was swept at 10 Oe/s in linear mode from $B_{max}$ to -$B_{max}$, then swept back to $B_{max}$. 
\end{itemize}

Loops in figure 4 were obtained by decreasing $B_{max}$. They are similar to experiments performed by increasing $B_{max}$. 

To measure local magnetization, we employed an array of Hall sensors based on high-mobility AlGaAs/GaAs heterostructure with a 160 nm two-dimensional electron gas (2DEG) below the surface, as done before \cite{li2019chiral,Collignon2017,Behnia2000}. The device was fabricated using electron beam lithography and 250 V argon ions to define the mesa. The device consists of ten  $5\times 5$ $\mu$m$^2$ sensors separated from their neighbor by 100 $\mu$m. The hall resistance of the sensors are $R_{Hall}\approx6.2 \times 10^3*B$. The local magnetic field at the surface of the sample was obtained by measuring the Hall resistivity of a sensor put on the sample. The sensor resistivity was measured using a Quantum Design PPMS with applied field along the c-axis of the sample.

\bibliography{Biblio}

\begin{thebibliography}{48}%
\makeatletter
\providecommand \@ifxundefined [1]{%
 \@ifx{#1\undefined}
}%
\providecommand \@ifnum [1]{%
 \ifnum #1\expandafter \@firstoftwo
 \else \expandafter \@secondoftwo
 \fi
}%
\providecommand \@ifx [1]{%
 \ifx #1\expandafter \@firstoftwo
 \else \expandafter \@secondoftwo
 \fi
}%
\providecommand \natexlab [1]{#1}%
\providecommand \enquote  [1]{``#1''}%
\providecommand \bibnamefont  [1]{#1}%
\providecommand \bibfnamefont [1]{#1}%
\providecommand \citenamefont [1]{#1}%
\providecommand \href@noop [0]{\@secondoftwo}%
\providecommand \href [0]{\begingroup \@sanitize@url \@href}%
\providecommand \@href[1]{\@@startlink{#1}\@@href}%
\providecommand \@@href[1]{\endgroup#1\@@endlink}%
\providecommand \@sanitize@url [0]{\catcode `\\12\catcode `\$12\catcode `\&12\catcode `\#12\catcode `\^12\catcode `\_12\catcode `\%12\relax}%
\providecommand \@@startlink[1]{}%
\providecommand \@@endlink[0]{}%
\providecommand \url  [0]{\begingroup\@sanitize@url \@url }%
\providecommand \@url [1]{\endgroup\@href {#1}{\urlprefix }}%
\providecommand \urlprefix  [0]{URL }%
\providecommand \Eprint [0]{\href }%
\providecommand \doibase [0]{https://doi.org/}%
\providecommand \selectlanguage [0]{\@gobble}%
\providecommand \bibinfo  [0]{\@secondoftwo}%
\providecommand \bibfield  [0]{\@secondoftwo}%
\providecommand \translation [1]{[#1]}%
\providecommand \BibitemOpen [0]{}%
\providecommand \bibitemStop [0]{}%
\providecommand \bibitemNoStop [0]{.\EOS\space}%
\providecommand \EOS [0]{\spacefactor3000\relax}%
\providecommand \BibitemShut  [1]{\csname bibitem#1\endcsname}%
\let\auto@bib@innerbib\@empty
\bibitem [{\citenamefont {Zabel}\ \emph {et~al.}(1979)\citenamefont {Zabel}, \citenamefont {Wandinger},\ and\ \citenamefont {Range}}]{ZabelWandingerRange+1979+238+241}%
  \BibitemOpen
  \bibfield  {author} {\bibinfo {author} {\bibfnamefont {M.}~\bibnamefont {Zabel}}, \bibinfo {author} {\bibfnamefont {S.}~\bibnamefont {Wandinger}},\ and\ \bibinfo {author} {\bibfnamefont {K.-J.}\ \bibnamefont {Range}},\ }\bibfield  {title} {\bibinfo {title} {{Ternary Chalcogenides M$_3$M$_2$X$_2$ with Shandite-Type Structure}},\ }\href {https://doi.org/doi:10.1515/znb-1979-0219} {\bibfield  {journal} {\bibinfo  {journal} {Zeitschrift für Naturforschung B}\ }\textbf {\bibinfo {volume} {34}},\ \bibinfo {pages} {238} (\bibinfo {year} {1979})}\BibitemShut {NoStop}%
\bibitem [{\citenamefont {Liu}\ \emph {et~al.}(2018)\citenamefont {Liu}, \citenamefont {Sun}, \citenamefont {Kumar}, \citenamefont {Muechler}, \citenamefont {Sun}, \citenamefont {Jiao}, \citenamefont {Yang}, \citenamefont {Liu}, \citenamefont {Liang}, \citenamefont {Xu} \emph {et~al.}}]{liu2018giant}%
  \BibitemOpen
  \bibfield  {author} {\bibinfo {author} {\bibfnamefont {E.}~\bibnamefont {Liu}}, \bibinfo {author} {\bibfnamefont {Y.}~\bibnamefont {Sun}}, \bibinfo {author} {\bibfnamefont {N.}~\bibnamefont {Kumar}}, \bibinfo {author} {\bibfnamefont {L.}~\bibnamefont {Muechler}}, \bibinfo {author} {\bibfnamefont {A.}~\bibnamefont {Sun}}, \bibinfo {author} {\bibfnamefont {L.}~\bibnamefont {Jiao}}, \bibinfo {author} {\bibfnamefont {S.-Y.}\ \bibnamefont {Yang}}, \bibinfo {author} {\bibfnamefont {D.}~\bibnamefont {Liu}}, \bibinfo {author} {\bibfnamefont {A.}~\bibnamefont {Liang}}, \bibinfo {author} {\bibfnamefont {Q.}~\bibnamefont {Xu}}, \emph {et~al.},\ }\bibfield  {title} {\bibinfo {title} {Giant anomalous {Hall} effect in a ferromagnetic kagome-lattice semimetal},\ }\href@noop {} {\bibfield  {journal} {\bibinfo  {journal} {Nature physics}\ }\textbf {\bibinfo {volume} {14}},\ \bibinfo {pages} {1125} (\bibinfo {year} {2018})}\BibitemShut {NoStop}%
\bibitem [{\citenamefont {Wang}\ \emph {et~al.}(2018)\citenamefont {Wang}, \citenamefont {Xu}, \citenamefont {Lou}, \citenamefont {Liu}, \citenamefont {Li}, \citenamefont {Huang}, \citenamefont {Shen}, \citenamefont {Weng}, \citenamefont {Wang},\ and\ \citenamefont {Lei}}]{wang2018large}%
  \BibitemOpen
  \bibfield  {author} {\bibinfo {author} {\bibfnamefont {Q.}~\bibnamefont {Wang}}, \bibinfo {author} {\bibfnamefont {Y.}~\bibnamefont {Xu}}, \bibinfo {author} {\bibfnamefont {R.}~\bibnamefont {Lou}}, \bibinfo {author} {\bibfnamefont {Z.}~\bibnamefont {Liu}}, \bibinfo {author} {\bibfnamefont {M.}~\bibnamefont {Li}}, \bibinfo {author} {\bibfnamefont {Y.}~\bibnamefont {Huang}}, \bibinfo {author} {\bibfnamefont {D.}~\bibnamefont {Shen}}, \bibinfo {author} {\bibfnamefont {H.}~\bibnamefont {Weng}}, \bibinfo {author} {\bibfnamefont {S.}~\bibnamefont {Wang}},\ and\ \bibinfo {author} {\bibfnamefont {H.}~\bibnamefont {Lei}},\ }\bibfield  {title} {\bibinfo {title} {Large intrinsic anomalous {Hall} effect in half-metallic ferromagnet {Co$_3$Sn$_2$S$_2$} with magnetic {Weyl} fermions},\ }\href@noop {} {\bibfield  {journal} {\bibinfo  {journal} {Nature communications}\ }\textbf {\bibinfo {volume} {9}},\ \bibinfo {pages} {3681} (\bibinfo {year} {2018})}\BibitemShut {NoStop}%
\bibitem [{\citenamefont {Vaqueiro}\ and\ \citenamefont {Sobany}(2009)}]{VAQUEIRO2009513}%
  \BibitemOpen
  \bibfield  {author} {\bibinfo {author} {\bibfnamefont {P.}~\bibnamefont {Vaqueiro}}\ and\ \bibinfo {author} {\bibfnamefont {G.~G.}\ \bibnamefont {Sobany}},\ }\bibfield  {title} {\bibinfo {title} {A powder neutron diffraction study of the metallic ferromagnet {Co$_3$Sn$_2$S$_2$}},\ }\href {https://doi.org/https://doi.org/10.1016/j.solidstatesciences.2008.06.017} {\bibfield  {journal} {\bibinfo  {journal} {Solid State Sciences}\ }\textbf {\bibinfo {volume} {11}},\ \bibinfo {pages} {513} (\bibinfo {year} {2009})}\BibitemShut {NoStop}%
\bibitem [{\citenamefont {Dedkov}\ \emph {et~al.}(2008)\citenamefont {Dedkov}, \citenamefont {Holder}, \citenamefont {Molodtsov},\ and\ \citenamefont {Rosner}}]{Dedkov_2008}%
  \BibitemOpen
  \bibfield  {author} {\bibinfo {author} {\bibfnamefont {Y.~S.}\ \bibnamefont {Dedkov}}, \bibinfo {author} {\bibfnamefont {M.}~\bibnamefont {Holder}}, \bibinfo {author} {\bibfnamefont {S.~L.}\ \bibnamefont {Molodtsov}},\ and\ \bibinfo {author} {\bibfnamefont {H.}~\bibnamefont {Rosner}},\ }\bibfield  {title} {\bibinfo {title} {Electronic structure of shandite {Co$_3$Sn$_2$S$_2$}},\ }\href {https://doi.org/10.1088/1742-6596/100/7/072011} {\bibfield  {journal} {\bibinfo  {journal} {Journal of Physics: Conference Series}\ }\textbf {\bibinfo {volume} {100}},\ \bibinfo {pages} {072011} (\bibinfo {year} {2008})}\BibitemShut {NoStop}%
\bibitem [{\citenamefont {Holder}\ \emph {et~al.}(2009)\citenamefont {Holder}, \citenamefont {Dedkov}, \citenamefont {Kade}, \citenamefont {Rosner}, \citenamefont {Schnelle}, \citenamefont {Leithe-Jasper}, \citenamefont {Weihrich},\ and\ \citenamefont {Molodtsov}}]{Holder2009}%
  \BibitemOpen
  \bibfield  {author} {\bibinfo {author} {\bibfnamefont {M.}~\bibnamefont {Holder}}, \bibinfo {author} {\bibfnamefont {Y.~S.}\ \bibnamefont {Dedkov}}, \bibinfo {author} {\bibfnamefont {A.}~\bibnamefont {Kade}}, \bibinfo {author} {\bibfnamefont {H.}~\bibnamefont {Rosner}}, \bibinfo {author} {\bibfnamefont {W.}~\bibnamefont {Schnelle}}, \bibinfo {author} {\bibfnamefont {A.}~\bibnamefont {Leithe-Jasper}}, \bibinfo {author} {\bibfnamefont {R.}~\bibnamefont {Weihrich}},\ and\ \bibinfo {author} {\bibfnamefont {S.~L.}\ \bibnamefont {Molodtsov}},\ }\bibfield  {title} {\bibinfo {title} {Photoemission study of electronic structure of the half-metallic ferromagnet {Co$_3$Sn$_2$S$_2$}},\ }\href {https://doi.org/10.1103/PhysRevB.79.205116} {\bibfield  {journal} {\bibinfo  {journal} {Phys. Rev. B}\ }\textbf {\bibinfo {volume} {79}},\ \bibinfo {pages} {205116} (\bibinfo {year} {2009})}\BibitemShut {NoStop}%
\bibitem [{\citenamefont {Schnelle}\ \emph {et~al.}(2013)\citenamefont {Schnelle}, \citenamefont {Leithe-Jasper}, \citenamefont {Rosner}, \citenamefont {Schappacher}, \citenamefont {P\"ottgen}, \citenamefont {Pielnhofer},\ and\ \citenamefont {Weihrich}}]{Schnelle2013}%
  \BibitemOpen
  \bibfield  {author} {\bibinfo {author} {\bibfnamefont {W.}~\bibnamefont {Schnelle}}, \bibinfo {author} {\bibfnamefont {A.}~\bibnamefont {Leithe-Jasper}}, \bibinfo {author} {\bibfnamefont {H.}~\bibnamefont {Rosner}}, \bibinfo {author} {\bibfnamefont {F.~M.}\ \bibnamefont {Schappacher}}, \bibinfo {author} {\bibfnamefont {R.}~\bibnamefont {P\"ottgen}}, \bibinfo {author} {\bibfnamefont {F.}~\bibnamefont {Pielnhofer}},\ and\ \bibinfo {author} {\bibfnamefont {R.}~\bibnamefont {Weihrich}},\ }\bibfield  {title} {\bibinfo {title} {Ferromagnetic ordering and half-metallic state of {Sn${}_{2}$Co${}_{3}$S${}_{2}$} with the shandite-type structure},\ }\href {https://doi.org/10.1103/PhysRevB.88.144404} {\bibfield  {journal} {\bibinfo  {journal} {Phys. Rev. B}\ }\textbf {\bibinfo {volume} {88}},\ \bibinfo {pages} {144404} (\bibinfo {year} {2013})}\BibitemShut {NoStop}%
\bibitem [{\citenamefont {Ding}\ \emph {et~al.}(2019)\citenamefont {Ding}, \citenamefont {Koo}, \citenamefont {Xu}, \citenamefont {Li}, \citenamefont {Lu}, \citenamefont {Zhao}, \citenamefont {Wang}, \citenamefont {Yin}, \citenamefont {Lei}, \citenamefont {Yan} \emph {et~al.}}]{ding2019intrinsic}%
  \BibitemOpen
  \bibfield  {author} {\bibinfo {author} {\bibfnamefont {L.}~\bibnamefont {Ding}}, \bibinfo {author} {\bibfnamefont {J.}~\bibnamefont {Koo}}, \bibinfo {author} {\bibfnamefont {L.}~\bibnamefont {Xu}}, \bibinfo {author} {\bibfnamefont {X.}~\bibnamefont {Li}}, \bibinfo {author} {\bibfnamefont {X.}~\bibnamefont {Lu}}, \bibinfo {author} {\bibfnamefont {L.}~\bibnamefont {Zhao}}, \bibinfo {author} {\bibfnamefont {Q.}~\bibnamefont {Wang}}, \bibinfo {author} {\bibfnamefont {Q.}~\bibnamefont {Yin}}, \bibinfo {author} {\bibfnamefont {H.}~\bibnamefont {Lei}}, \bibinfo {author} {\bibfnamefont {B.}~\bibnamefont {Yan}}, \emph {et~al.},\ }\bibfield  {title} {\bibinfo {title} {Intrinsic anomalous {Nernst} effect amplified by disorder in a half-metallic semimetal},\ }\href@noop {} {\bibfield  {journal} {\bibinfo  {journal} {Physical Review X}\ }\textbf {\bibinfo {volume} {9}},\ \bibinfo {pages} {041061} (\bibinfo {year} {2019})}\BibitemShut {NoStop}%
\bibitem [{\citenamefont {Ding}\ \emph {et~al.}(2021)\citenamefont {Ding}, \citenamefont {Koo}, \citenamefont {Yi}, \citenamefont {Xu}, \citenamefont {Zuo}, \citenamefont {Yang}, \citenamefont {Shi}, \citenamefont {Yan}, \citenamefont {Behnia},\ and\ \citenamefont {Zhu}}]{Ding_2021}%
  \BibitemOpen
  \bibfield  {author} {\bibinfo {author} {\bibfnamefont {L.}~\bibnamefont {Ding}}, \bibinfo {author} {\bibfnamefont {J.}~\bibnamefont {Koo}}, \bibinfo {author} {\bibfnamefont {C.}~\bibnamefont {Yi}}, \bibinfo {author} {\bibfnamefont {L.}~\bibnamefont {Xu}}, \bibinfo {author} {\bibfnamefont {H.}~\bibnamefont {Zuo}}, \bibinfo {author} {\bibfnamefont {M.}~\bibnamefont {Yang}}, \bibinfo {author} {\bibfnamefont {Y.}~\bibnamefont {Shi}}, \bibinfo {author} {\bibfnamefont {B.}~\bibnamefont {Yan}}, \bibinfo {author} {\bibfnamefont {K.}~\bibnamefont {Behnia}},\ and\ \bibinfo {author} {\bibfnamefont {Z.}~\bibnamefont {Zhu}},\ }\bibfield  {title} {\bibinfo {title} {Quantum oscillations, magnetic breakdown and thermal {Hall} effect in {Co$_3$Sn$_2$S$_2$}},\ }\href {https://doi.org/10.1088/1361-6463/ac1c2b} {\bibfield  {journal} {\bibinfo  {journal} {Journal of Physics D: Applied Physics}\ }\textbf {\bibinfo {volume} {54}},\ \bibinfo {pages} {454003} (\bibinfo {year} {2021})}\BibitemShut {NoStop}%
\bibitem [{\citenamefont {Sakai}\ \emph {et~al.}(2018)\citenamefont {Sakai}, \citenamefont {Mizuta}, \citenamefont {Nugroho}, \citenamefont {Sihombing}, \citenamefont {Koretsune}, \citenamefont {Suzuki}, \citenamefont {Takemori}, \citenamefont {Ishii}, \citenamefont {Nishio-Hamane}, \citenamefont {Arita}, \citenamefont {Goswami},\ and\ \citenamefont {Nakatsuji}}]{Sakai2018}%
  \BibitemOpen
  \bibfield  {author} {\bibinfo {author} {\bibfnamefont {A.}~\bibnamefont {Sakai}}, \bibinfo {author} {\bibfnamefont {Y.~P.}\ \bibnamefont {Mizuta}}, \bibinfo {author} {\bibfnamefont {A.~A.}\ \bibnamefont {Nugroho}}, \bibinfo {author} {\bibfnamefont {R.}~\bibnamefont {Sihombing}}, \bibinfo {author} {\bibfnamefont {T.}~\bibnamefont {Koretsune}}, \bibinfo {author} {\bibfnamefont {M.-T.}\ \bibnamefont {Suzuki}}, \bibinfo {author} {\bibfnamefont {N.}~\bibnamefont {Takemori}}, \bibinfo {author} {\bibfnamefont {R.}~\bibnamefont {Ishii}}, \bibinfo {author} {\bibfnamefont {D.}~\bibnamefont {Nishio-Hamane}}, \bibinfo {author} {\bibfnamefont {R.}~\bibnamefont {Arita}}, \bibinfo {author} {\bibfnamefont {P.}~\bibnamefont {Goswami}},\ and\ \bibinfo {author} {\bibfnamefont {S.}~\bibnamefont {Nakatsuji}},\ }\bibfield  {title} {\bibinfo {title} {{Giant anomalous Nernst effect and quantum-critical scaling in a ferromagnetic semimetal}},\ }\href {https://doi.org/10.1038/s41567-018-0225-6} {\bibfield  {journal} {\bibinfo
  {journal} {Nature Physics}\ }\textbf {\bibinfo {volume} {14}},\ \bibinfo {pages} {1119} (\bibinfo {year} {2018})}\BibitemShut {NoStop}%
\bibitem [{\citenamefont {Li}\ \emph {et~al.}(2017)\citenamefont {Li}, \citenamefont {Xu}, \citenamefont {Ding}, \citenamefont {Wang}, \citenamefont {Shen}, \citenamefont {Lu}, \citenamefont {Zhu},\ and\ \citenamefont {Behnia}}]{Li2017}%
  \BibitemOpen
  \bibfield  {author} {\bibinfo {author} {\bibfnamefont {X.}~\bibnamefont {Li}}, \bibinfo {author} {\bibfnamefont {L.}~\bibnamefont {Xu}}, \bibinfo {author} {\bibfnamefont {L.}~\bibnamefont {Ding}}, \bibinfo {author} {\bibfnamefont {J.}~\bibnamefont {Wang}}, \bibinfo {author} {\bibfnamefont {M.}~\bibnamefont {Shen}}, \bibinfo {author} {\bibfnamefont {X.}~\bibnamefont {Lu}}, \bibinfo {author} {\bibfnamefont {Z.}~\bibnamefont {Zhu}},\ and\ \bibinfo {author} {\bibfnamefont {K.}~\bibnamefont {Behnia}},\ }\bibfield  {title} {\bibinfo {title} {Anomalous {Nernst and Righi-Leduc} effects in {${\mathrm{Mn}}_{3}\mathrm{Sn}$: Berry} curvature and entropy flow},\ }\href {https://doi.org/10.1103/PhysRevLett.119.056601} {\bibfield  {journal} {\bibinfo  {journal} {Phys. Rev. Lett.}\ }\textbf {\bibinfo {volume} {119}},\ \bibinfo {pages} {056601} (\bibinfo {year} {2017})}\BibitemShut {NoStop}%
\bibitem [{\citenamefont {Ikhlas}\ \emph {et~al.}(2017)\citenamefont {Ikhlas}, \citenamefont {Tomita}, \citenamefont {Koretsune}, \citenamefont {Suzuki}, \citenamefont {Nishio-Hamane}, \citenamefont {Arita}, \citenamefont {Otani},\ and\ \citenamefont {Nakatsuji}}]{Ikhlas2017}%
  \BibitemOpen
  \bibfield  {author} {\bibinfo {author} {\bibfnamefont {M.}~\bibnamefont {Ikhlas}}, \bibinfo {author} {\bibfnamefont {T.}~\bibnamefont {Tomita}}, \bibinfo {author} {\bibfnamefont {T.}~\bibnamefont {Koretsune}}, \bibinfo {author} {\bibfnamefont {M.-T.}\ \bibnamefont {Suzuki}}, \bibinfo {author} {\bibfnamefont {D.}~\bibnamefont {Nishio-Hamane}}, \bibinfo {author} {\bibfnamefont {R.}~\bibnamefont {Arita}}, \bibinfo {author} {\bibfnamefont {Y.}~\bibnamefont {Otani}},\ and\ \bibinfo {author} {\bibfnamefont {S.}~\bibnamefont {Nakatsuji}},\ }\bibfield  {title} {\bibinfo {title} {{Large anomalous Nernst effect at room temperature in a chiral antiferromagnet}},\ }\href {https://doi.org/10.1038/NPHYS4181} {\bibfield  {journal} {\bibinfo  {journal} {Nature Physics}\ }\textbf {\bibinfo {volume} {13}},\ \bibinfo {pages} {1085} (\bibinfo {year} {2017})}\BibitemShut {NoStop}%
\bibitem [{\citenamefont {Xu}\ \emph {et~al.}(2020{\natexlab{a}})\citenamefont {Xu}, \citenamefont {Li}, \citenamefont {Lu}, \citenamefont {Collignon}, \citenamefont {Fu}, \citenamefont {Koo}, \citenamefont {Fauqué}, \citenamefont {Yan}, \citenamefont {Zhu},\ and\ \citenamefont {Behnia}}]{xu2020-2}%
  \BibitemOpen
  \bibfield  {author} {\bibinfo {author} {\bibfnamefont {L.}~\bibnamefont {Xu}}, \bibinfo {author} {\bibfnamefont {X.}~\bibnamefont {Li}}, \bibinfo {author} {\bibfnamefont {X.}~\bibnamefont {Lu}}, \bibinfo {author} {\bibfnamefont {C.}~\bibnamefont {Collignon}}, \bibinfo {author} {\bibfnamefont {H.}~\bibnamefont {Fu}}, \bibinfo {author} {\bibfnamefont {J.}~\bibnamefont {Koo}}, \bibinfo {author} {\bibfnamefont {B.}~\bibnamefont {Fauqué}}, \bibinfo {author} {\bibfnamefont {B.}~\bibnamefont {Yan}}, \bibinfo {author} {\bibfnamefont {Z.}~\bibnamefont {Zhu}},\ and\ \bibinfo {author} {\bibfnamefont {K.}~\bibnamefont {Behnia}},\ }\bibfield  {title} {\bibinfo {title} {Finite-temperature violation of the anomalous transverse {Wiedemann-Franz} law},\ }\href {https://doi.org/10.1126/sciadv.aaz3522} {\bibfield  {journal} {\bibinfo  {journal} {Science Advances}\ }\textbf {\bibinfo {volume} {6}},\ \bibinfo {pages} {eaaz3522} (\bibinfo {year} {2020}{\natexlab{a}})}\BibitemShut {NoStop}%
\bibitem [{\citenamefont {Ren}\ \emph {et~al.}(2022)\citenamefont {Ren}, \citenamefont {Li}, \citenamefont {Sharma}, \citenamefont {Bhattarai}, \citenamefont {Zhao}, \citenamefont {Rachmilowitz}, \citenamefont {Bahrami}, \citenamefont {Tafti}, \citenamefont {Fang}, \citenamefont {Ghimire}, \citenamefont {Wang},\ and\ \citenamefont {Zeljkovic}}]{Ren2022}%
  \BibitemOpen
  \bibfield  {author} {\bibinfo {author} {\bibfnamefont {Z.}~\bibnamefont {Ren}}, \bibinfo {author} {\bibfnamefont {H.}~\bibnamefont {Li}}, \bibinfo {author} {\bibfnamefont {S.}~\bibnamefont {Sharma}}, \bibinfo {author} {\bibfnamefont {D.}~\bibnamefont {Bhattarai}}, \bibinfo {author} {\bibfnamefont {H.}~\bibnamefont {Zhao}}, \bibinfo {author} {\bibfnamefont {B.}~\bibnamefont {Rachmilowitz}}, \bibinfo {author} {\bibfnamefont {F.}~\bibnamefont {Bahrami}}, \bibinfo {author} {\bibfnamefont {F.}~\bibnamefont {Tafti}}, \bibinfo {author} {\bibfnamefont {S.}~\bibnamefont {Fang}}, \bibinfo {author} {\bibfnamefont {M.~P.}\ \bibnamefont {Ghimire}}, \bibinfo {author} {\bibfnamefont {Z.}~\bibnamefont {Wang}},\ and\ \bibinfo {author} {\bibfnamefont {I.}~\bibnamefont {Zeljkovic}},\ }\bibfield  {title} {\bibinfo {title} {Plethora of tunable weyl fermions in kagome magnet {Fe$_3$Sn$_2$} thin films},\ }\href {https://doi.org/10.1038/s41535-022-00521-y} {\bibfield  {journal} {\bibinfo  {journal} {npj Quantum Materials}\
  }\textbf {\bibinfo {volume} {7}},\ \bibinfo {pages} {109} (\bibinfo {year} {2022})}\BibitemShut {NoStop}%
\bibitem [{\citenamefont {Guguchia}\ \emph {et~al.}(2020)\citenamefont {Guguchia}, \citenamefont {Verezhak}, \citenamefont {Gawryluk}, \citenamefont {Tsirkin}, \citenamefont {Yin}, \citenamefont {Belopolski}, \citenamefont {Zhou}, \citenamefont {Simutis}, \citenamefont {Zhang}, \citenamefont {Cochran} \emph {et~al.}}]{guguchia2020tunable}%
  \BibitemOpen
  \bibfield  {author} {\bibinfo {author} {\bibfnamefont {Z.}~\bibnamefont {Guguchia}}, \bibinfo {author} {\bibfnamefont {J.}~\bibnamefont {Verezhak}}, \bibinfo {author} {\bibfnamefont {D.}~\bibnamefont {Gawryluk}}, \bibinfo {author} {\bibfnamefont {S.}~\bibnamefont {Tsirkin}}, \bibinfo {author} {\bibfnamefont {J.-X.}\ \bibnamefont {Yin}}, \bibinfo {author} {\bibfnamefont {I.}~\bibnamefont {Belopolski}}, \bibinfo {author} {\bibfnamefont {H.}~\bibnamefont {Zhou}}, \bibinfo {author} {\bibfnamefont {G.}~\bibnamefont {Simutis}}, \bibinfo {author} {\bibfnamefont {S.-S.}\ \bibnamefont {Zhang}}, \bibinfo {author} {\bibfnamefont {T.}~\bibnamefont {Cochran}}, \emph {et~al.},\ }\bibfield  {title} {\bibinfo {title} {Tunable anomalous hall conductivity through volume-wise magnetic competition in a topological kagome magnet},\ }\href@noop {} {\bibfield  {journal} {\bibinfo  {journal} {Nature communications}\ }\textbf {\bibinfo {volume} {11}},\ \bibinfo {pages} {559} (\bibinfo {year} {2020})}\BibitemShut {NoStop}%
\bibitem [{\citenamefont {Zhang}\ \emph {et~al.}(2021)\citenamefont {Zhang}, \citenamefont {Okamoto}, \citenamefont {Samolyuk}, \citenamefont {Stone}, \citenamefont {Kolesnikov}, \citenamefont {Xue}, \citenamefont {Yan}, \citenamefont {McGuire}, \citenamefont {Mandrus},\ and\ \citenamefont {Tennant}}]{zhang2021unusual}%
  \BibitemOpen
  \bibfield  {author} {\bibinfo {author} {\bibfnamefont {Q.}~\bibnamefont {Zhang}}, \bibinfo {author} {\bibfnamefont {S.}~\bibnamefont {Okamoto}}, \bibinfo {author} {\bibfnamefont {G.~D.}\ \bibnamefont {Samolyuk}}, \bibinfo {author} {\bibfnamefont {M.~B.}\ \bibnamefont {Stone}}, \bibinfo {author} {\bibfnamefont {A.~I.}\ \bibnamefont {Kolesnikov}}, \bibinfo {author} {\bibfnamefont {R.}~\bibnamefont {Xue}}, \bibinfo {author} {\bibfnamefont {J.}~\bibnamefont {Yan}}, \bibinfo {author} {\bibfnamefont {M.~A.}\ \bibnamefont {McGuire}}, \bibinfo {author} {\bibfnamefont {D.}~\bibnamefont {Mandrus}},\ and\ \bibinfo {author} {\bibfnamefont {D.~A.}\ \bibnamefont {Tennant}},\ }\bibfield  {title} {\bibinfo {title} {{Unusual Exchange Couplings and Intermediate Temperature Weyl State in Co$_3$Sn$_2$S$_2$}},\ }\href@noop {} {\bibfield  {journal} {\bibinfo  {journal} {Physical Review Letters}\ }\textbf {\bibinfo {volume} {127}},\ \bibinfo {pages} {117201} (\bibinfo {year} {2021})}\BibitemShut {NoStop}%
\bibitem [{\citenamefont {Lee}\ \emph {et~al.}(2022)\citenamefont {Lee}, \citenamefont {Vir}, \citenamefont {Manna}, \citenamefont {Shekhar}, \citenamefont {Moore}, \citenamefont {Kastner}, \citenamefont {Felser},\ and\ \citenamefont {Orenstein}}]{lee2022observation}%
  \BibitemOpen
  \bibfield  {author} {\bibinfo {author} {\bibfnamefont {C.}~\bibnamefont {Lee}}, \bibinfo {author} {\bibfnamefont {P.}~\bibnamefont {Vir}}, \bibinfo {author} {\bibfnamefont {K.}~\bibnamefont {Manna}}, \bibinfo {author} {\bibfnamefont {C.}~\bibnamefont {Shekhar}}, \bibinfo {author} {\bibfnamefont {J.}~\bibnamefont {Moore}}, \bibinfo {author} {\bibfnamefont {M.}~\bibnamefont {Kastner}}, \bibinfo {author} {\bibfnamefont {C.}~\bibnamefont {Felser}},\ and\ \bibinfo {author} {\bibfnamefont {J.}~\bibnamefont {Orenstein}},\ }\bibfield  {title} {\bibinfo {title} {Observation of a phase transition within the domain walls of ferromagnetic {Co$_3$Sn$_2$S$_2$}},\ }\href@noop {} {\bibfield  {journal} {\bibinfo  {journal} {Nature communications}\ }\textbf {\bibinfo {volume} {13}},\ \bibinfo {pages} {3000} (\bibinfo {year} {2022})}\BibitemShut {NoStop}%
\bibitem [{\citenamefont {Soh}\ \emph {et~al.}(2022)\citenamefont {Soh}, \citenamefont {Yi}, \citenamefont {Zivkovic}, \citenamefont {Qureshi}, \citenamefont {Stunault}, \citenamefont {Ouladdiaf}, \citenamefont {Rodr{\'\i}guez-Velamaz{\'a}n}, \citenamefont {Shi}, \citenamefont {R{\o}nnow},\ and\ \citenamefont {Boothroyd}}]{soh2022magnetic}%
  \BibitemOpen
  \bibfield  {author} {\bibinfo {author} {\bibfnamefont {J.-R.}\ \bibnamefont {Soh}}, \bibinfo {author} {\bibfnamefont {C.}~\bibnamefont {Yi}}, \bibinfo {author} {\bibfnamefont {I.}~\bibnamefont {Zivkovic}}, \bibinfo {author} {\bibfnamefont {N.}~\bibnamefont {Qureshi}}, \bibinfo {author} {\bibfnamefont {A.}~\bibnamefont {Stunault}}, \bibinfo {author} {\bibfnamefont {B.}~\bibnamefont {Ouladdiaf}}, \bibinfo {author} {\bibfnamefont {J.~A.}\ \bibnamefont {Rodr{\'\i}guez-Velamaz{\'a}n}}, \bibinfo {author} {\bibfnamefont {Y.}~\bibnamefont {Shi}}, \bibinfo {author} {\bibfnamefont {H.~M.}\ \bibnamefont {R{\o}nnow}},\ and\ \bibinfo {author} {\bibfnamefont {A.~T.}\ \bibnamefont {Boothroyd}},\ }\bibfield  {title} {\bibinfo {title} {Magnetic structure of the topological semimetal {Co$_3$Sn$_2$S$_2$}},\ }\href@noop {} {\bibfield  {journal} {\bibinfo  {journal} {Physical Review B}\ }\textbf {\bibinfo {volume} {105}},\ \bibinfo {pages} {094435} (\bibinfo {year} {2022})}\BibitemShut {NoStop}%
\bibitem [{\citenamefont {Neubauer}\ \emph {et~al.}(2022)\citenamefont {Neubauer}, \citenamefont {Ye}, \citenamefont {Shi}, \citenamefont {Malinowski}, \citenamefont {Gao}, \citenamefont {Taddei}, \citenamefont {Bourges}, \citenamefont {Ivanov}, \citenamefont {Chu},\ and\ \citenamefont {Dai}}]{neubauer2022spin}%
  \BibitemOpen
  \bibfield  {author} {\bibinfo {author} {\bibfnamefont {K.~J.}\ \bibnamefont {Neubauer}}, \bibinfo {author} {\bibfnamefont {F.}~\bibnamefont {Ye}}, \bibinfo {author} {\bibfnamefont {Y.}~\bibnamefont {Shi}}, \bibinfo {author} {\bibfnamefont {P.}~\bibnamefont {Malinowski}}, \bibinfo {author} {\bibfnamefont {B.}~\bibnamefont {Gao}}, \bibinfo {author} {\bibfnamefont {K.~M.}\ \bibnamefont {Taddei}}, \bibinfo {author} {\bibfnamefont {P.}~\bibnamefont {Bourges}}, \bibinfo {author} {\bibfnamefont {A.}~\bibnamefont {Ivanov}}, \bibinfo {author} {\bibfnamefont {J.-H.}\ \bibnamefont {Chu}},\ and\ \bibinfo {author} {\bibfnamefont {P.}~\bibnamefont {Dai}},\ }\bibfield  {title} {\bibinfo {title} {{Spin structure and dynamics of the topological semimetal Co$_3$Sn$_{2-x}$In $_x$ S$_2$}},\ }\href@noop {} {\bibfield  {journal} {\bibinfo  {journal} {npj Quantum Materials}\ }\textbf {\bibinfo {volume} {7}},\ \bibinfo {pages} {112} (\bibinfo {year} {2022})}\BibitemShut {NoStop}%
\bibitem [{\citenamefont {Kassem}\ \emph {et~al.}(2017)\citenamefont {Kassem}, \citenamefont {Tabata}, \citenamefont {Waki},\ and\ \citenamefont {Nakamura}}]{kassem2017low}%
  \BibitemOpen
  \bibfield  {author} {\bibinfo {author} {\bibfnamefont {M.~A.}\ \bibnamefont {Kassem}}, \bibinfo {author} {\bibfnamefont {Y.}~\bibnamefont {Tabata}}, \bibinfo {author} {\bibfnamefont {T.}~\bibnamefont {Waki}},\ and\ \bibinfo {author} {\bibfnamefont {H.}~\bibnamefont {Nakamura}},\ }\bibfield  {title} {\bibinfo {title} {Low-field anomalous magnetic phase in the kagome-lattice shandite {Co$_3$Sn$_2$S$_2$}},\ }\href@noop {} {\bibfield  {journal} {\bibinfo  {journal} {Physical Review B}\ }\textbf {\bibinfo {volume} {96}},\ \bibinfo {pages} {014429} (\bibinfo {year} {2017})}\BibitemShut {NoStop}%
\bibitem [{\citenamefont {Lachman}\ \emph {et~al.}(2020)\citenamefont {Lachman}, \citenamefont {Murphy}, \citenamefont {Maksimovic}, \citenamefont {Kealhofer}, \citenamefont {Haley}, \citenamefont {McDonald}, \citenamefont {Long},\ and\ \citenamefont {Analytis}}]{lachman2020exchange}%
  \BibitemOpen
  \bibfield  {author} {\bibinfo {author} {\bibfnamefont {E.}~\bibnamefont {Lachman}}, \bibinfo {author} {\bibfnamefont {R.~A.}\ \bibnamefont {Murphy}}, \bibinfo {author} {\bibfnamefont {N.}~\bibnamefont {Maksimovic}}, \bibinfo {author} {\bibfnamefont {R.}~\bibnamefont {Kealhofer}}, \bibinfo {author} {\bibfnamefont {S.}~\bibnamefont {Haley}}, \bibinfo {author} {\bibfnamefont {R.~D.}\ \bibnamefont {McDonald}}, \bibinfo {author} {\bibfnamefont {J.~R.}\ \bibnamefont {Long}},\ and\ \bibinfo {author} {\bibfnamefont {J.~G.}\ \bibnamefont {Analytis}},\ }\bibfield  {title} {\bibinfo {title} {Exchange biased anomalous {Hall} effect driven by frustration in a magnetic kagome lattice},\ }\href@noop {} {\bibfield  {journal} {\bibinfo  {journal} {Nature communications}\ }\textbf {\bibinfo {volume} {11}},\ \bibinfo {pages} {560} (\bibinfo {year} {2020})}\BibitemShut {NoStop}%
\bibitem [{\citenamefont {{\v{Z}}ivkovi{\'c}}\ \emph {et~al.}(2022)\citenamefont {{\v{Z}}ivkovi{\'c}}, \citenamefont {Yadav}, \citenamefont {Soh}, \citenamefont {Yi}, \citenamefont {Shi}, \citenamefont {Yazyev},\ and\ \citenamefont {R{\o}nnow}}]{vzivkovic2022unraveling}%
  \BibitemOpen
  \bibfield  {author} {\bibinfo {author} {\bibfnamefont {I.}~\bibnamefont {{\v{Z}}ivkovi{\'c}}}, \bibinfo {author} {\bibfnamefont {R.}~\bibnamefont {Yadav}}, \bibinfo {author} {\bibfnamefont {J.-R.}\ \bibnamefont {Soh}}, \bibinfo {author} {\bibfnamefont {C.}~\bibnamefont {Yi}}, \bibinfo {author} {\bibfnamefont {Y.}~\bibnamefont {Shi}}, \bibinfo {author} {\bibfnamefont {O.~V.}\ \bibnamefont {Yazyev}},\ and\ \bibinfo {author} {\bibfnamefont {H.~M.}\ \bibnamefont {R{\o}nnow}},\ }\bibfield  {title} {\bibinfo {title} {{Unraveling the origin of the peculiar transition in the magnetically ordered phase of the Weyl semimetal Co$_3$Sn$_2$S$_2$}},\ }\href@noop {} {\bibfield  {journal} {\bibinfo  {journal} {Physical Review B}\ }\textbf {\bibinfo {volume} {106}},\ \bibinfo {pages} {L180403} (\bibinfo {year} {2022})}\BibitemShut {NoStop}%
\bibitem [{\citenamefont {Avakyants}\ \emph {et~al.}(2023)\citenamefont {Avakyants}, \citenamefont {Orlova}, \citenamefont {Timonina}, \citenamefont {Kolesnikov},\ and\ \citenamefont {Deviatov}}]{avakyants2023evidence}%
  \BibitemOpen
  \bibfield  {author} {\bibinfo {author} {\bibfnamefont {A.}~\bibnamefont {Avakyants}}, \bibinfo {author} {\bibfnamefont {N.}~\bibnamefont {Orlova}}, \bibinfo {author} {\bibfnamefont {A.}~\bibnamefont {Timonina}}, \bibinfo {author} {\bibfnamefont {N.}~\bibnamefont {Kolesnikov}},\ and\ \bibinfo {author} {\bibfnamefont {E.}~\bibnamefont {Deviatov}},\ }\bibfield  {title} {\bibinfo {title} {Evidence for surface spin structures from first order reversal curves in {Co$_3$Sn$_2$S$_2$} and {Fe$_3$GeTe$_2$} magnetic topological semimetals},\ }\href@noop {} {\bibfield  {journal} {\bibinfo  {journal} {Journal of Magnetism and Magnetic Materials}\ }\textbf {\bibinfo {volume} {573}},\ \bibinfo {pages} {170668} (\bibinfo {year} {2023})}\BibitemShut {NoStop}%
\bibitem [{\citenamefont {Noah}\ \emph {et~al.}(2022)\citenamefont {Noah}, \citenamefont {Toric}, \citenamefont {Feld}, \citenamefont {Zissman}, \citenamefont {Gutfreund}, \citenamefont {Tsruya}, \citenamefont {Devidas}, \citenamefont {Alpern}, \citenamefont {Vakahi}, \citenamefont {Steinberg} \emph {et~al.}}]{noah2022tunable}%
  \BibitemOpen
  \bibfield  {author} {\bibinfo {author} {\bibfnamefont {A.}~\bibnamefont {Noah}}, \bibinfo {author} {\bibfnamefont {F.}~\bibnamefont {Toric}}, \bibinfo {author} {\bibfnamefont {T.~D.}\ \bibnamefont {Feld}}, \bibinfo {author} {\bibfnamefont {G.}~\bibnamefont {Zissman}}, \bibinfo {author} {\bibfnamefont {A.}~\bibnamefont {Gutfreund}}, \bibinfo {author} {\bibfnamefont {D.}~\bibnamefont {Tsruya}}, \bibinfo {author} {\bibfnamefont {T.}~\bibnamefont {Devidas}}, \bibinfo {author} {\bibfnamefont {H.}~\bibnamefont {Alpern}}, \bibinfo {author} {\bibfnamefont {A.}~\bibnamefont {Vakahi}}, \bibinfo {author} {\bibfnamefont {H.}~\bibnamefont {Steinberg}}, \emph {et~al.},\ }\bibfield  {title} {\bibinfo {title} {{Tunable exchange bias in the magnetic Weyl semimetal Co$_3$Sn$_2$S$_2$}},\ }\href@noop {} {\bibfield  {journal} {\bibinfo  {journal} {Physical Review B}\ }\textbf {\bibinfo {volume} {105}},\ \bibinfo {pages} {144423} (\bibinfo {year} {2022})}\BibitemShut {NoStop}%
\bibitem [{\citenamefont {Shen}\ \emph {et~al.}(2022)\citenamefont {Shen}, \citenamefont {Zhu}, \citenamefont {Ullah}, \citenamefont {Klavins},\ and\ \citenamefont {Taufour}}]{shen2022anomalous}%
  \BibitemOpen
  \bibfield  {author} {\bibinfo {author} {\bibfnamefont {Z.}~\bibnamefont {Shen}}, \bibinfo {author} {\bibfnamefont {X.}~\bibnamefont {Zhu}}, \bibinfo {author} {\bibfnamefont {R.~R.}\ \bibnamefont {Ullah}}, \bibinfo {author} {\bibfnamefont {P.}~\bibnamefont {Klavins}},\ and\ \bibinfo {author} {\bibfnamefont {V.}~\bibnamefont {Taufour}},\ }\bibfield  {title} {\bibinfo {title} {{Anomalous depinning of magnetic domain walls within the ferromagnetic phase of the Weyl semimetal Co$_3$Sn$_2$S$_2$}},\ }\href@noop {} {\bibfield  {journal} {\bibinfo  {journal} {Journal of Physics: Condensed Matter}\ }\textbf {\bibinfo {volume} {35}},\ \bibinfo {pages} {045802} (\bibinfo {year} {2022})}\BibitemShut {NoStop}%
\bibitem [{\citenamefont {Zhang}\ \emph {et~al.}(2022)\citenamefont {Zhang}, \citenamefont {Zhang}, \citenamefont {Matsuda}, \citenamefont {Garlea}, \citenamefont {Yan}, \citenamefont {McGuire}, \citenamefont {Tennant},\ and\ \citenamefont {Okamoto}}]{zhang2022hidden}%
  \BibitemOpen
  \bibfield  {author} {\bibinfo {author} {\bibfnamefont {Q.}~\bibnamefont {Zhang}}, \bibinfo {author} {\bibfnamefont {Y.}~\bibnamefont {Zhang}}, \bibinfo {author} {\bibfnamefont {M.}~\bibnamefont {Matsuda}}, \bibinfo {author} {\bibfnamefont {V.~O.}\ \bibnamefont {Garlea}}, \bibinfo {author} {\bibfnamefont {J.}~\bibnamefont {Yan}}, \bibinfo {author} {\bibfnamefont {M.~A.}\ \bibnamefont {McGuire}}, \bibinfo {author} {\bibfnamefont {D.~A.}\ \bibnamefont {Tennant}},\ and\ \bibinfo {author} {\bibfnamefont {S.}~\bibnamefont {Okamoto}},\ }\bibfield  {title} {\bibinfo {title} {{Hidden local symmetry breaking in a kagome-lattice magnetic Weyl semimetal}},\ }\href@noop {} {\bibfield  {journal} {\bibinfo  {journal} {Journal of the American Chemical Society}\ }\textbf {\bibinfo {volume} {144}},\ \bibinfo {pages} {14339} (\bibinfo {year} {2022})}\BibitemShut {NoStop}%
\bibitem [{\citenamefont {Stamps}(2000)}]{Stamps_2000}%
  \BibitemOpen
  \bibfield  {author} {\bibinfo {author} {\bibfnamefont {R.~L.}\ \bibnamefont {Stamps}},\ }\bibfield  {title} {\bibinfo {title} {Mechanisms for exchange bias},\ }\href {https://doi.org/10.1088/0022-3727/33/23/201} {\bibfield  {journal} {\bibinfo  {journal} {Journal of Physics D: Applied Physics}\ }\textbf {\bibinfo {volume} {33}},\ \bibinfo {pages} {R247} (\bibinfo {year} {2000})}\BibitemShut {NoStop}%
\bibitem [{\citenamefont {Pate}\ \emph {et~al.}(2023)\citenamefont {Pate}, \citenamefont {Wang}, \citenamefont {Shen}, \citenamefont {Jiang}, \citenamefont {Welp}, \citenamefont {Kwok}, \citenamefont {Xu}, \citenamefont {Li}, \citenamefont {Divan},\ and\ \citenamefont {Xiao}}]{pate2023field}%
  \BibitemOpen
  \bibfield  {author} {\bibinfo {author} {\bibfnamefont {S.~E.}\ \bibnamefont {Pate}}, \bibinfo {author} {\bibfnamefont {B.}~\bibnamefont {Wang}}, \bibinfo {author} {\bibfnamefont {B.}~\bibnamefont {Shen}}, \bibinfo {author} {\bibfnamefont {J.~S.}\ \bibnamefont {Jiang}}, \bibinfo {author} {\bibfnamefont {U.}~\bibnamefont {Welp}}, \bibinfo {author} {\bibfnamefont {W.-K.}\ \bibnamefont {Kwok}}, \bibinfo {author} {\bibfnamefont {J.}~\bibnamefont {Xu}}, \bibinfo {author} {\bibfnamefont {K.}~\bibnamefont {Li}}, \bibinfo {author} {\bibfnamefont {R.}~\bibnamefont {Divan}},\ and\ \bibinfo {author} {\bibfnamefont {Z.-L.}\ \bibnamefont {Xiao}},\ }\bibfield  {title} {\bibinfo {title} {Field orientation dependent magnetic phases in the weyl semimetal {Co$_3$Sn$_2$S$_2$}},\ }\href@noop {} {\bibfield  {journal} {\bibinfo  {journal} {Physical Review B}\ }\textbf {\bibinfo {volume} {108}},\ \bibinfo {pages} {L100408} (\bibinfo {year} {2023})}\BibitemShut {NoStop}%
\bibitem [{\citenamefont {Keim}\ \emph {et~al.}(2019)\citenamefont {Keim}, \citenamefont {Paulsen}, \citenamefont {Zeravcic}, \citenamefont {Sastry},\ and\ \citenamefont {Nagel}}]{Keim2019}%
  \BibitemOpen
  \bibfield  {author} {\bibinfo {author} {\bibfnamefont {N.~C.}\ \bibnamefont {Keim}}, \bibinfo {author} {\bibfnamefont {J.~D.}\ \bibnamefont {Paulsen}}, \bibinfo {author} {\bibfnamefont {Z.}~\bibnamefont {Zeravcic}}, \bibinfo {author} {\bibfnamefont {S.}~\bibnamefont {Sastry}},\ and\ \bibinfo {author} {\bibfnamefont {S.~R.}\ \bibnamefont {Nagel}},\ }\bibfield  {title} {\bibinfo {title} {Memory formation in matter},\ }\href {https://doi.org/10.1103/RevModPhys.91.035002} {\bibfield  {journal} {\bibinfo  {journal} {Rev. Mod. Phys.}\ }\textbf {\bibinfo {volume} {91}},\ \bibinfo {pages} {035002} (\bibinfo {year} {2019})}\BibitemShut {NoStop}%
\bibitem [{\citenamefont {Li}\ \emph {et~al.}(2019)\citenamefont {Li}, \citenamefont {Collignon}, \citenamefont {Xu}, \citenamefont {Zuo}, \citenamefont {Cavanna}, \citenamefont {Gennser}, \citenamefont {Mailly}, \citenamefont {Fauqu{\'e}}, \citenamefont {Balents}, \citenamefont {Zhu} \emph {et~al.}}]{li2019chiral}%
  \BibitemOpen
  \bibfield  {author} {\bibinfo {author} {\bibfnamefont {X.}~\bibnamefont {Li}}, \bibinfo {author} {\bibfnamefont {C.}~\bibnamefont {Collignon}}, \bibinfo {author} {\bibfnamefont {L.}~\bibnamefont {Xu}}, \bibinfo {author} {\bibfnamefont {H.}~\bibnamefont {Zuo}}, \bibinfo {author} {\bibfnamefont {A.}~\bibnamefont {Cavanna}}, \bibinfo {author} {\bibfnamefont {U.}~\bibnamefont {Gennser}}, \bibinfo {author} {\bibfnamefont {D.}~\bibnamefont {Mailly}}, \bibinfo {author} {\bibfnamefont {B.}~\bibnamefont {Fauqu{\'e}}}, \bibinfo {author} {\bibfnamefont {L.}~\bibnamefont {Balents}}, \bibinfo {author} {\bibfnamefont {Z.}~\bibnamefont {Zhu}}, \emph {et~al.},\ }\bibfield  {title} {\bibinfo {title} {Chiral domain walls of {Mn$_3$Sn} and their memory},\ }\href@noop {} {\bibfield  {journal} {\bibinfo  {journal} {Nature communications}\ }\textbf {\bibinfo {volume} {10}},\ \bibinfo {pages} {3021} (\bibinfo {year} {2019})}\BibitemShut {NoStop}%
\bibitem [{\citenamefont {Xu}\ \emph {et~al.}(2020{\natexlab{b}})\citenamefont {Xu}, \citenamefont {Li}, \citenamefont {Ding}, \citenamefont {Behnia},\ and\ \citenamefont {Zhu}}]{Xu2020}%
  \BibitemOpen
  \bibfield  {author} {\bibinfo {author} {\bibfnamefont {L.}~\bibnamefont {Xu}}, \bibinfo {author} {\bibfnamefont {X.}~\bibnamefont {Li}}, \bibinfo {author} {\bibfnamefont {L.}~\bibnamefont {Ding}}, \bibinfo {author} {\bibfnamefont {K.}~\bibnamefont {Behnia}},\ and\ \bibinfo {author} {\bibfnamefont {Z.}~\bibnamefont {Zhu}},\ }\bibfield  {title} {\bibinfo {title} {{Planar Hall effect caused by the memory of antiferromagnetic domain walls in Mn$_3$Ge}},\ }\href {https://doi.org/10.1063/5.0030546} {\bibfield  {journal} {\bibinfo  {journal} {Applied Physics Letters}\ }\textbf {\bibinfo {volume} {117}},\ \bibinfo {pages} {222403} (\bibinfo {year} {2020}{\natexlab{b}})}\BibitemShut {NoStop}%
\bibitem [{\citenamefont {Sugawara}\ \emph {et~al.}(2019)\citenamefont {Sugawara}, \citenamefont {Akashi}, \citenamefont {Kassem}, \citenamefont {Tabata}, \citenamefont {Waki},\ and\ \citenamefont {Nakamura}}]{sugawara2019magnetic}%
  \BibitemOpen
  \bibfield  {author} {\bibinfo {author} {\bibfnamefont {A.}~\bibnamefont {Sugawara}}, \bibinfo {author} {\bibfnamefont {T.}~\bibnamefont {Akashi}}, \bibinfo {author} {\bibfnamefont {M.~A.}\ \bibnamefont {Kassem}}, \bibinfo {author} {\bibfnamefont {Y.}~\bibnamefont {Tabata}}, \bibinfo {author} {\bibfnamefont {T.}~\bibnamefont {Waki}},\ and\ \bibinfo {author} {\bibfnamefont {H.}~\bibnamefont {Nakamura}},\ }\bibfield  {title} {\bibinfo {title} {Magnetic domain structure within half-metallic ferromagnetic kagome compound {Co$_3$Sn$_2$S$_2$}},\ }\href@noop {} {\bibfield  {journal} {\bibinfo  {journal} {Physical Review Materials}\ }\textbf {\bibinfo {volume} {3}},\ \bibinfo {pages} {104421} (\bibinfo {year} {2019})}\BibitemShut {NoStop}%
\bibitem [{\citenamefont {Howlader}\ \emph {et~al.}(2020)\citenamefont {Howlader}, \citenamefont {Ramachandran}, \citenamefont {Singh}, \citenamefont {Sheet} \emph {et~al.}}]{howlader2020domain}%
  \BibitemOpen
  \bibfield  {author} {\bibinfo {author} {\bibfnamefont {S.}~\bibnamefont {Howlader}}, \bibinfo {author} {\bibfnamefont {R.}~\bibnamefont {Ramachandran}}, \bibinfo {author} {\bibfnamefont {Y.}~\bibnamefont {Singh}}, \bibinfo {author} {\bibfnamefont {G.}~\bibnamefont {Sheet}}, \emph {et~al.},\ }\bibfield  {title} {\bibinfo {title} {{Domain structure evolution in the ferromagnetic Kagome-lattice Weyl semimetal Co$_3$Sn$_2$S$_2$}},\ }\href@noop {} {\bibfield  {journal} {\bibinfo  {journal} {Journal of Physics: Condensed Matter}\ }\textbf {\bibinfo {volume} {33}},\ \bibinfo {pages} {075801} (\bibinfo {year} {2020})}\BibitemShut {NoStop}%
\bibitem [{\citenamefont {Aharoni}(1998)}]{aharoni1998demagnetizing}%
  \BibitemOpen
  \bibfield  {author} {\bibinfo {author} {\bibfnamefont {A.}~\bibnamefont {Aharoni}},\ }\bibfield  {title} {\bibinfo {title} {Demagnetizing factors for rectangular ferromagnetic prisms},\ }\href@noop {} {\bibfield  {journal} {\bibinfo  {journal} {Journal of applied physics}\ }\textbf {\bibinfo {volume} {83}},\ \bibinfo {pages} {3432} (\bibinfo {year} {1998})}\BibitemShut {NoStop}%
\bibitem [{\citenamefont {Garc{\i}a-Otero}\ \emph {et~al.}(1998)\citenamefont {Garc{\i}a-Otero}, \citenamefont {Garc{\i}a-Bastida},\ and\ \citenamefont {Rivas}}]{garcia1998influence}%
  \BibitemOpen
  \bibfield  {author} {\bibinfo {author} {\bibfnamefont {J.}~\bibnamefont {Garc{\i}a-Otero}}, \bibinfo {author} {\bibfnamefont {A.}~\bibnamefont {Garc{\i}a-Bastida}},\ and\ \bibinfo {author} {\bibfnamefont {J.}~\bibnamefont {Rivas}},\ }\bibfield  {title} {\bibinfo {title} {Influence of temperature on the coercive field of non-interacting fine magnetic particles},\ }\href@noop {} {\bibfield  {journal} {\bibinfo  {journal} {Journal of magnetism and magnetic materials}\ }\textbf {\bibinfo {volume} {189}},\ \bibinfo {pages} {377} (\bibinfo {year} {1998})}\BibitemShut {NoStop}%
\bibitem [{\citenamefont {Shen}\ \emph {et~al.}(2019)\citenamefont {Shen}, \citenamefont {Zeng}, \citenamefont {Zhang}, \citenamefont {Tong}, \citenamefont {Ling}, \citenamefont {Xi}, \citenamefont {Wang}, \citenamefont {Liu}, \citenamefont {Wang}, \citenamefont {Wu} \emph {et~al.}}]{shen2019anisotropies}%
  \BibitemOpen
  \bibfield  {author} {\bibinfo {author} {\bibfnamefont {J.}~\bibnamefont {Shen}}, \bibinfo {author} {\bibfnamefont {Q.}~\bibnamefont {Zeng}}, \bibinfo {author} {\bibfnamefont {S.}~\bibnamefont {Zhang}}, \bibinfo {author} {\bibfnamefont {W.}~\bibnamefont {Tong}}, \bibinfo {author} {\bibfnamefont {L.}~\bibnamefont {Ling}}, \bibinfo {author} {\bibfnamefont {C.}~\bibnamefont {Xi}}, \bibinfo {author} {\bibfnamefont {Z.}~\bibnamefont {Wang}}, \bibinfo {author} {\bibfnamefont {E.}~\bibnamefont {Liu}}, \bibinfo {author} {\bibfnamefont {W.}~\bibnamefont {Wang}}, \bibinfo {author} {\bibfnamefont {G.}~\bibnamefont {Wu}}, \emph {et~al.},\ }\bibfield  {title} {\bibinfo {title} {{On the anisotropies of magnetization and electronic transport of magnetic Weyl semimetal Co$_3$Sn$_2$S$_2$}},\ }\href@noop {} {\bibfield  {journal} {\bibinfo  {journal} {Applied Physics Letters}\ }\textbf {\bibinfo {volume} {115}} (\bibinfo {year} {2019})}\BibitemShut {NoStop}%
\bibitem [{\citenamefont {Hartmann}(1987)}]{hartmann1987origin}%
  \BibitemOpen
  \bibfield  {author} {\bibinfo {author} {\bibfnamefont {U.}~\bibnamefont {Hartmann}},\ }\bibfield  {title} {\bibinfo {title} {Origin of {Brown's} coercive paradox in perfect ferromagnetic crystals},\ }\href@noop {} {\bibfield  {journal} {\bibinfo  {journal} {Physical Review B}\ }\textbf {\bibinfo {volume} {36}},\ \bibinfo {pages} {2331} (\bibinfo {year} {1987})}\BibitemShut {NoStop}%
\bibitem [{\citenamefont {Coey}(2010)}]{coey2010magnetism}%
  \BibitemOpen
  \bibfield  {author} {\bibinfo {author} {\bibfnamefont {J.~M.}\ \bibnamefont {Coey}},\ }\href@noop {} {\emph {\bibinfo {title} {Magnetism and magnetic materials}}}\ (\bibinfo  {publisher} {Cambridge university press},\ \bibinfo {year} {2010})\ p.\ \bibinfo {pages} {245}\BibitemShut {NoStop}%
\bibitem [{\citenamefont {Aharoni}(1960)}]{aharoni1960reduction}%
  \BibitemOpen
  \bibfield  {author} {\bibinfo {author} {\bibfnamefont {A.}~\bibnamefont {Aharoni}},\ }\bibfield  {title} {\bibinfo {title} {Reduction in coercive force caused by a certain type of imperfection},\ }\href@noop {} {\bibfield  {journal} {\bibinfo  {journal} {Physical Review}\ }\textbf {\bibinfo {volume} {119}},\ \bibinfo {pages} {127} (\bibinfo {year} {1960})}\BibitemShut {NoStop}%
\bibitem [{SM(2024)}]{SM}%
  \BibitemOpen
  \href@noop {} {}\bibinfo {howpublished} {See {Supplemental Material} for more details} (\bibinfo {year} {2024})\BibitemShut {NoStop}%
\bibitem [{\citenamefont {Wales}(2001)}]{Wales2001}%
  \BibitemOpen
  \bibfield  {author} {\bibinfo {author} {\bibfnamefont {D.~J.}\ \bibnamefont {Wales}},\ }\bibinfo {title} {Energy landscapes},\ in\ \href@noop {} {\emph {\bibinfo {booktitle} {Atomic clusters and nanoparticles. Agregats atomiques et nanoparticules: Les Houches Session LXXIII 2--28 July 2000}}},\ \bibinfo {editor} {edited by\ \bibinfo {editor} {\bibfnamefont {C.}~\bibnamefont {Guet}}, \bibinfo {editor} {\bibfnamefont {P.}~\bibnamefont {Hobza}}, \bibinfo {editor} {\bibfnamefont {F.}~\bibnamefont {Speigelman}},\ and\ \bibinfo {editor} {\bibfnamefont {F.}~\bibnamefont {David}}}\ (\bibinfo  {publisher} {Springer Berlin Heidelberg},\ \bibinfo {address} {Berlin, Heidelberg},\ \bibinfo {year} {2001})\ pp.\ \bibinfo {pages} {437--507}\BibitemShut {NoStop}%
\bibitem [{\citenamefont {Hill}(1994)}]{hill1994thermodynamics}%
  \BibitemOpen
  \bibfield  {author} {\bibinfo {author} {\bibfnamefont {T.}~\bibnamefont {Hill}},\ }\href {https://books.google.fr/books?id=Xa-yAAAAQBAJ} {\emph {\bibinfo {title} {Thermodynamics of Small Systems}}},\ Dover Books on Chemistry\ (\bibinfo  {publisher} {Dover Publications},\ \bibinfo {year} {1994})\BibitemShut {NoStop}%
\bibitem [{\citenamefont {Ding}\ \emph {et~al.}(2023)\citenamefont {Ding}, \citenamefont {Chen}, \citenamefont {Wang}, \citenamefont {Zhang}, \citenamefont {Yang}, \citenamefont {Bi}, \citenamefont {Lin}, \citenamefont {Wang}, \citenamefont {Wu}, \citenamefont {Gu}, \citenamefont {Meng}, \citenamefont {Cao}, \citenamefont {Gu}, \citenamefont {Zhang}, \citenamefont {Zhong}, \citenamefont {Liu},\ and\ \citenamefont {Guo}}]{Ding2023}%
  \BibitemOpen
  \bibfield  {author} {\bibinfo {author} {\bibfnamefont {Z.}~\bibnamefont {Ding}}, \bibinfo {author} {\bibfnamefont {X.}~\bibnamefont {Chen}}, \bibinfo {author} {\bibfnamefont {Z.}~\bibnamefont {Wang}}, \bibinfo {author} {\bibfnamefont {Q.}~\bibnamefont {Zhang}}, \bibinfo {author} {\bibfnamefont {F.}~\bibnamefont {Yang}}, \bibinfo {author} {\bibfnamefont {J.}~\bibnamefont {Bi}}, \bibinfo {author} {\bibfnamefont {T.}~\bibnamefont {Lin}}, \bibinfo {author} {\bibfnamefont {Z.}~\bibnamefont {Wang}}, \bibinfo {author} {\bibfnamefont {X.}~\bibnamefont {Wu}}, \bibinfo {author} {\bibfnamefont {M.}~\bibnamefont {Gu}}, \bibinfo {author} {\bibfnamefont {M.}~\bibnamefont {Meng}}, \bibinfo {author} {\bibfnamefont {Y.}~\bibnamefont {Cao}}, \bibinfo {author} {\bibfnamefont {L.}~\bibnamefont {Gu}}, \bibinfo {author} {\bibfnamefont {J.}~\bibnamefont {Zhang}}, \bibinfo {author} {\bibfnamefont {Z.}~\bibnamefont {Zhong}}, \bibinfo {author} {\bibfnamefont {X.}~\bibnamefont {Liu}},\ and\ \bibinfo {author} {\bibfnamefont
  {J.}~\bibnamefont {Guo}},\ }\bibfield  {title} {\bibinfo {title} {Magnetism and berry phase manipulation in an emergent structure of perovskite ruthenate by (111) strain engineering},\ }\href {https://doi.org/10.1038/s41535-023-00576-5} {\bibfield  {journal} {\bibinfo  {journal} {npj Quantum Materials}\ }\textbf {\bibinfo {volume} {8}},\ \bibinfo {pages} {43} (\bibinfo {year} {2023})}\BibitemShut {NoStop}%
\bibitem [{\citenamefont {Kar}\ \emph {et~al.}(2023)\citenamefont {Kar}, \citenamefont {Singh}, \citenamefont {Hsu}, \citenamefont {Lin}, \citenamefont {Das}, \citenamefont {Cheng}, \citenamefont {Berben}, \citenamefont {Yang}, \citenamefont {Lin}, \citenamefont {Hsu}, \citenamefont {Wiedmann}, \citenamefont {Lee},\ and\ \citenamefont {Lee}}]{Kar2023}%
  \BibitemOpen
  \bibfield  {author} {\bibinfo {author} {\bibfnamefont {U.}~\bibnamefont {Kar}}, \bibinfo {author} {\bibfnamefont {A.~K.}\ \bibnamefont {Singh}}, \bibinfo {author} {\bibfnamefont {Y.-T.}\ \bibnamefont {Hsu}}, \bibinfo {author} {\bibfnamefont {C.-Y.}\ \bibnamefont {Lin}}, \bibinfo {author} {\bibfnamefont {B.}~\bibnamefont {Das}}, \bibinfo {author} {\bibfnamefont {C.-T.}\ \bibnamefont {Cheng}}, \bibinfo {author} {\bibfnamefont {M.}~\bibnamefont {Berben}}, \bibinfo {author} {\bibfnamefont {S.}~\bibnamefont {Yang}}, \bibinfo {author} {\bibfnamefont {C.-Y.}\ \bibnamefont {Lin}}, \bibinfo {author} {\bibfnamefont {C.-H.}\ \bibnamefont {Hsu}}, \bibinfo {author} {\bibfnamefont {S.}~\bibnamefont {Wiedmann}}, \bibinfo {author} {\bibfnamefont {W.-C.}\ \bibnamefont {Lee}},\ and\ \bibinfo {author} {\bibfnamefont {W.-L.}\ \bibnamefont {Lee}},\ }\bibfield  {title} {\bibinfo {title} {{The thickness dependence of quantum oscillations in ferromagnetic Weyl metal SrRuO$_3$}},\ }\href {https://doi.org/10.1038/s41535-023-00540-3}
  {\bibfield  {journal} {\bibinfo  {journal} {npj Quantum Materials}\ }\textbf {\bibinfo {volume} {8}},\ \bibinfo {pages} {8} (\bibinfo {year} {2023})}\BibitemShut {NoStop}%
\bibitem [{\citenamefont {Kimbell}\ \emph {et~al.}(2020)\citenamefont {Kimbell}, \citenamefont {Sass}, \citenamefont {Woltjes}, \citenamefont {Ko}, \citenamefont {Noh}, \citenamefont {Wu},\ and\ \citenamefont {Robinson}}]{kimbell2020two}%
  \BibitemOpen
  \bibfield  {author} {\bibinfo {author} {\bibfnamefont {G.}~\bibnamefont {Kimbell}}, \bibinfo {author} {\bibfnamefont {P.~M.}\ \bibnamefont {Sass}}, \bibinfo {author} {\bibfnamefont {B.}~\bibnamefont {Woltjes}}, \bibinfo {author} {\bibfnamefont {E.~K.}\ \bibnamefont {Ko}}, \bibinfo {author} {\bibfnamefont {T.~W.}\ \bibnamefont {Noh}}, \bibinfo {author} {\bibfnamefont {W.}~\bibnamefont {Wu}},\ and\ \bibinfo {author} {\bibfnamefont {J.~W.}\ \bibnamefont {Robinson}},\ }\bibfield  {title} {\bibinfo {title} {{Two-channel anomalous Hall effect in SrRuO$_3$}},\ }\href@noop {} {\bibfield  {journal} {\bibinfo  {journal} {Physical Review Materials}\ }\textbf {\bibinfo {volume} {4}},\ \bibinfo {pages} {054414} (\bibinfo {year} {2020})}\BibitemShut {NoStop}%
\bibitem [{\citenamefont {Zhou}\ \emph {et~al.}(2023)\citenamefont {Zhou}, \citenamefont {Kang}, \citenamefont {Ji}, \citenamefont {Dou}, \citenamefont {Wang},\ and\ \citenamefont {Xu}}]{zhou2023observation}%
  \BibitemOpen
  \bibfield  {author} {\bibinfo {author} {\bibfnamefont {G.}~\bibnamefont {Zhou}}, \bibinfo {author} {\bibfnamefont {P.}~\bibnamefont {Kang}}, \bibinfo {author} {\bibfnamefont {H.}~\bibnamefont {Ji}}, \bibinfo {author} {\bibfnamefont {J.}~\bibnamefont {Dou}}, \bibinfo {author} {\bibfnamefont {S.}~\bibnamefont {Wang}},\ and\ \bibinfo {author} {\bibfnamefont {X.}~\bibnamefont {Xu}},\ }\bibfield  {title} {\bibinfo {title} {{Observation of two-channel anomalous Hall effect in perpendicularly magnetized NiCo$_2$O$_4$ epitaxial films}},\ }\href@noop {} {\bibfield  {journal} {\bibinfo  {journal} {Physical Review B}\ }\textbf {\bibinfo {volume} {108}},\ \bibinfo {pages} {094442} (\bibinfo {year} {2023})}\BibitemShut {NoStop}%
\bibitem [{\citenamefont {Collignon}\ \emph {et~al.}(2017)\citenamefont {Collignon}, \citenamefont {Fauqu\'e}, \citenamefont {Cavanna}, \citenamefont {Gennser}, \citenamefont {Mailly},\ and\ \citenamefont {Behnia}}]{Collignon2017}%
  \BibitemOpen
  \bibfield  {author} {\bibinfo {author} {\bibfnamefont {C.}~\bibnamefont {Collignon}}, \bibinfo {author} {\bibfnamefont {B.}~\bibnamefont {Fauqu\'e}}, \bibinfo {author} {\bibfnamefont {A.}~\bibnamefont {Cavanna}}, \bibinfo {author} {\bibfnamefont {U.}~\bibnamefont {Gennser}}, \bibinfo {author} {\bibfnamefont {D.}~\bibnamefont {Mailly}},\ and\ \bibinfo {author} {\bibfnamefont {K.}~\bibnamefont {Behnia}},\ }\bibfield  {title} {\bibinfo {title} {Superfluid density and carrier concentration across a superconducting dome: The case of strontium titanate},\ }\href {https://doi.org/10.1103/PhysRevB.96.224506} {\bibfield  {journal} {\bibinfo  {journal} {Phys. Rev. B}\ }\textbf {\bibinfo {volume} {96}},\ \bibinfo {pages} {224506} (\bibinfo {year} {2017})}\BibitemShut {NoStop}%
\bibitem [{\citenamefont {Behnia}\ \emph {et~al.}(2000)\citenamefont {Behnia}, \citenamefont {Capan}, \citenamefont {Mailly},\ and\ \citenamefont {Etienne}}]{Behnia2000}%
  \BibitemOpen
  \bibfield  {author} {\bibinfo {author} {\bibfnamefont {K.}~\bibnamefont {Behnia}}, \bibinfo {author} {\bibfnamefont {C.}~\bibnamefont {Capan}}, \bibinfo {author} {\bibfnamefont {D.}~\bibnamefont {Mailly}},\ and\ \bibinfo {author} {\bibfnamefont {B.}~\bibnamefont {Etienne}},\ }\bibfield  {title} {\bibinfo {title} {Internal avalanches in a pile of superconducting vortices},\ }\href {https://doi.org/10.1103/PhysRevB.61.R3815} {\bibfield  {journal} {\bibinfo  {journal} {Phys. Rev. B}\ }\textbf {\bibinfo {volume} {61}},\ \bibinfo {pages} {R3815} (\bibinfo {year} {2000})}\BibitemShut {NoStop}%
\end{thebibliography}%

\newpage

\begin{center}{\large\bf Supplementary Materials for ``Magnetic memory and distinct spin populations in ferromagnetic Co$_{3}$Sn$_{2}$S$_{2}$"}\\
\end{center}

\renewcommand{\thesection}{Supplementary note \arabic{section}}
\renewcommand{\thetable}{Supplementary \arabic{table}}
\renewcommand{\thefigure}{Supplementary \arabic{figure}}
\renewcommand{\theequation}{Supplementary \arabic{equation}}

\setcounter{section}{0}
\setcounter{figure}{0}
\setcounter{table}{0}
\setcounter{equation}{0}

\section{Samples}

\begin{figure}[ht]
    \centering
    \includegraphics[width=0.4\textwidth]{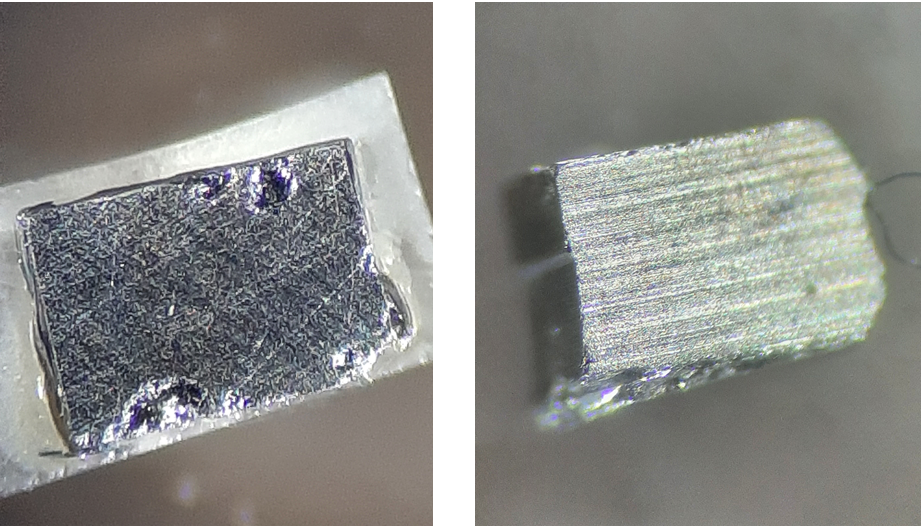}
\caption{\label{fig:supp cristaux}\justifying{\small{Picture of Co$_3$Sn$_2$S$_2$ samples. (left) Sample 1, (right) Sample 2.}}}
\end{figure}
Supplementary figure 1 shows pictures of Co$_{3}$Sn$_{2}$S$_{2}$ crystals used in this study. Sample dimensions are 1.21$\times$0.89$\times$0.14 mm$^3$ for sample 1 and 1.93$\times$1.20$\times$1.38 mm$^3$ for sample 2.

\section{Resistivity data}

\begin{figure}[ht]
    \centering
    \includegraphics[width=0.45\textwidth]{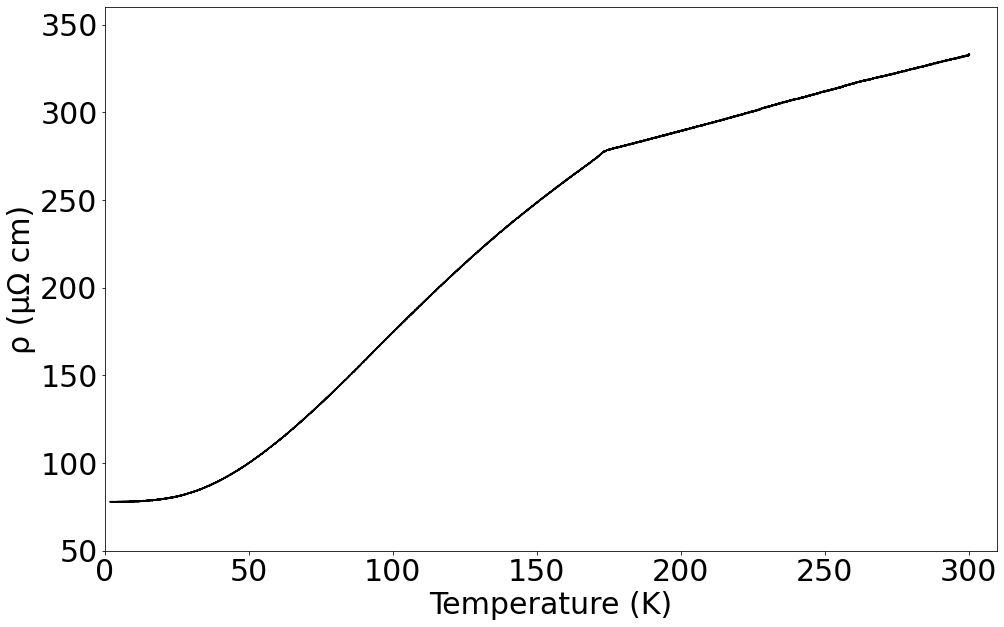}
\caption{\justifying{\small{Resistivity of sample 1 versus temperature at zero field.}}}
\end{figure}
\begin{figure}[ht]
    \centering
    \includegraphics[width=0.45\textwidth]{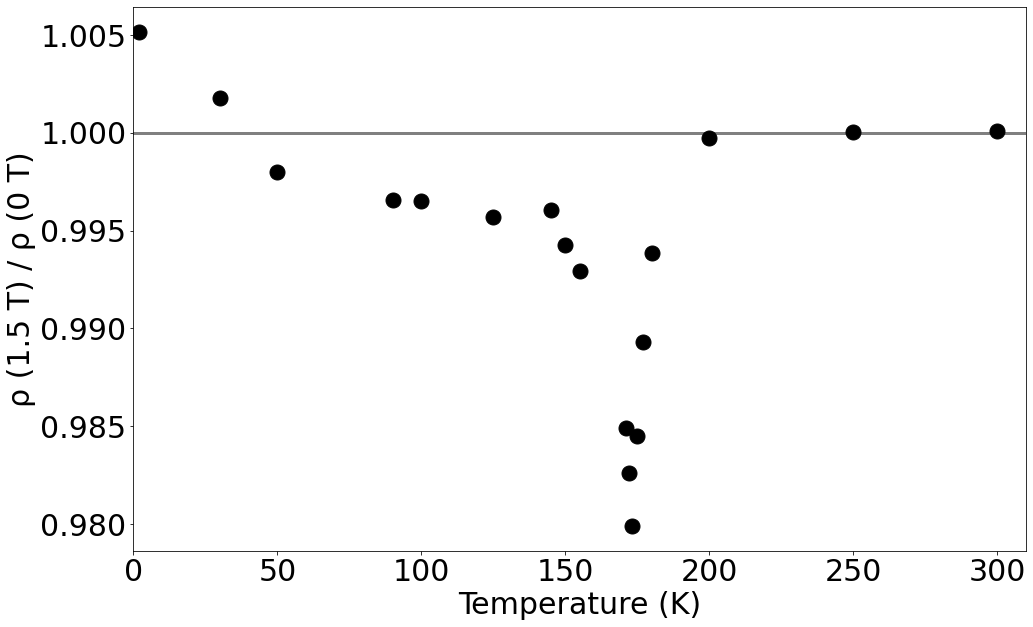}
\caption{\justifying{\small{Resistivity at 1.5 T normalized by resistivity at zero field as a function of temperature.}}}
\end{figure}
Supplementary figure 2 displays resistivity of sample 1 versus temperature. We observe a kink at 173 K, corresponding to the ferromagnetic-paramagnetic transition at Curie temperature $T_C$. 

Supplementary figure 3 displays resistivity under an applied field of 1.5 T normalised by resistivity at zero field. There is a minimum around 173 K, which also correspond to $T_C$.

\section{Magnetization measurements}
Performing multiple loops with increasing $B_{max}$ causes $B_0$ to increase with some steps, which may vary between measurements. Supplementary figure 4 compares $B_{0+}$ measurement for increasing and decreasing $B_{max}$. There is no significant difference between increasing and decreasing $B_{max}$. Both protocols gave similar results.

\begin{figure}[ht]
\centering
    \includegraphics[width=0.45\textwidth]{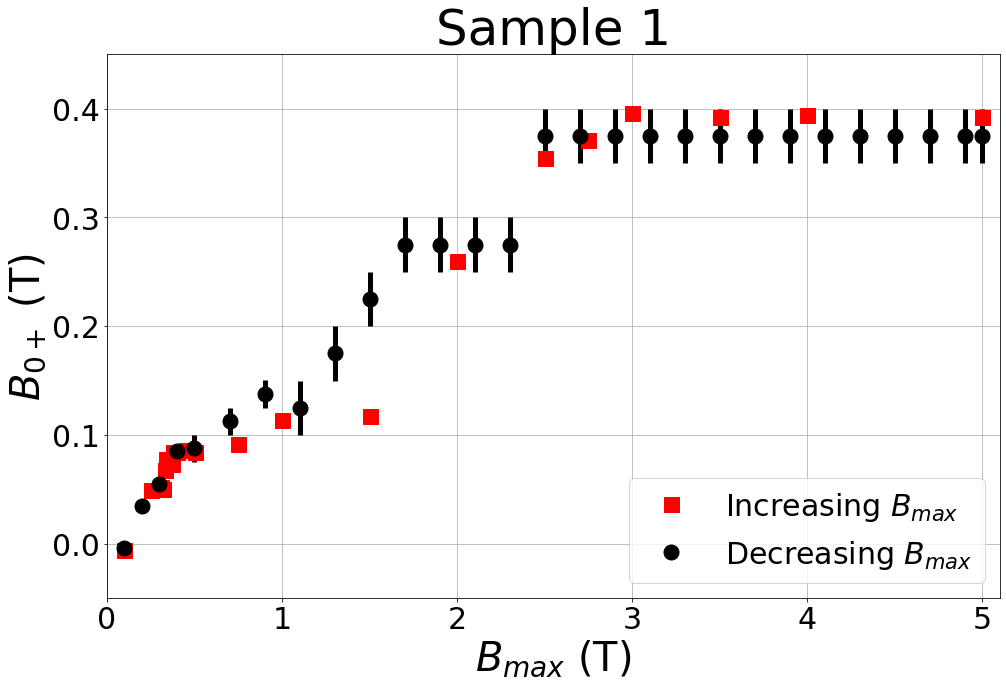}
\caption{\label{fig:Bc_plus}\justifying{\small{Positive spin flip field ($B_{0+}$ ) for increasing and decreasing $B_{max}$.}}}
\end{figure}

\section{Exchange Bias}

\begin{figure}[ht]
\centering
    \includegraphics[width=0.45\textwidth]{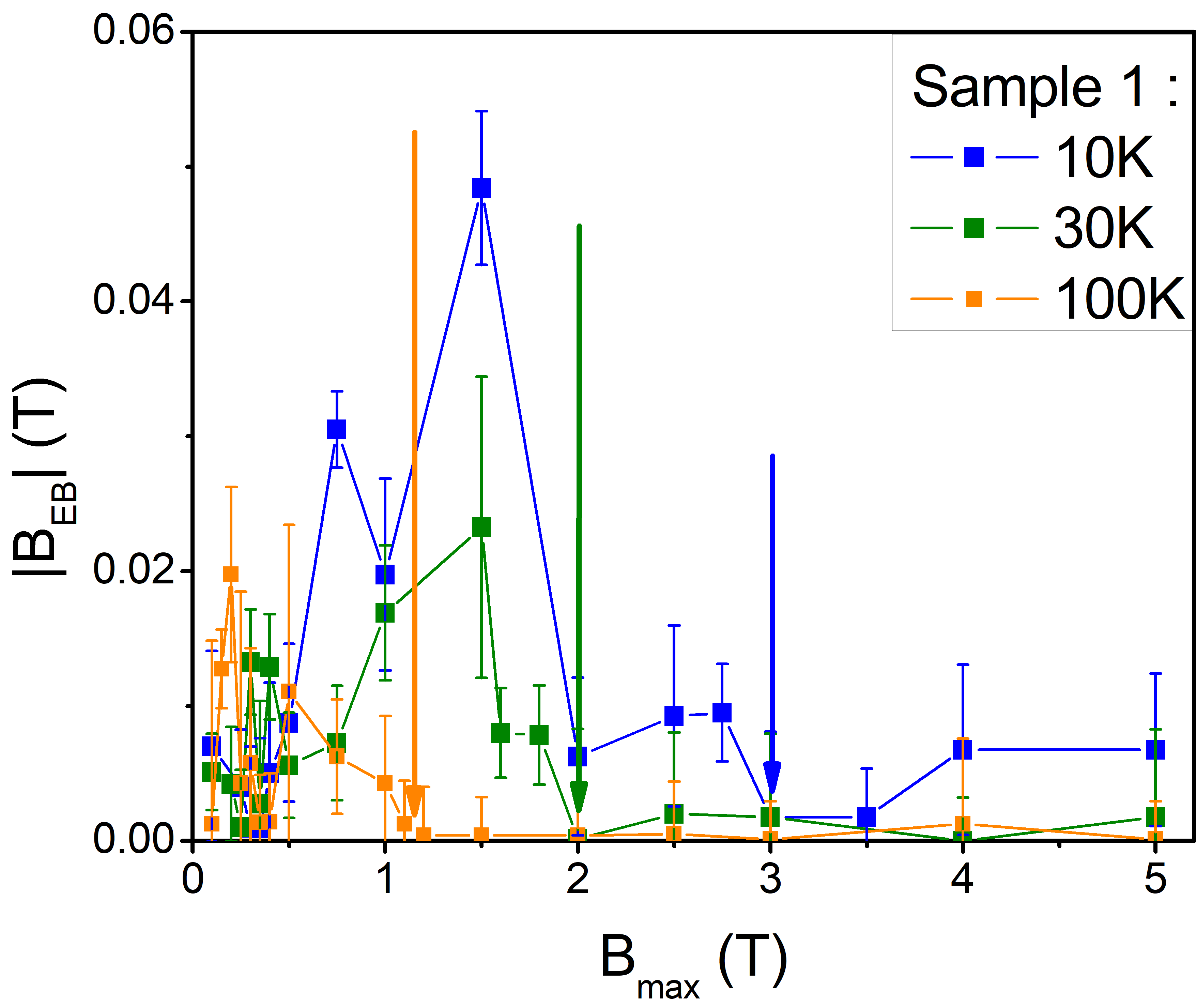}
\caption{\label{fig:Beb_Bmax}\justifying{\small{Exchange bias field, $|B_{EB}|$ as a function of B$_{max}$ at several temperatures.}}}
\end{figure}

The exchange bias field, defined as $B_{EB}=(|B_{0+}|-|B_{0-}|)/2$ depends on $B_{max}$ too. Supplementary figure 5 shows $B_{EB}$ at 10 K, 30 K and 100 K. The amplitude of $B_{EB}$ was found to strongly fluctuates and its sign differs between successive hysteresis loops. However, when B$_{max}$>B$_{max}^\star$, the amplitude of $B_{EB}$ becomes as small as the statistical error margin.

\newpage

\section{Local magnetization}
Supplementary figure 6 is a picture of the arrays of Hall sensors used to measure local magnetization of our samples. he device consists of ten  $5\times 5$ $\mu$m$^2$ sensors separated from their neighbor by 100 $\mu$m.

\begin{figure}[ht]
\centering
    \includegraphics[width=0.45\textwidth]{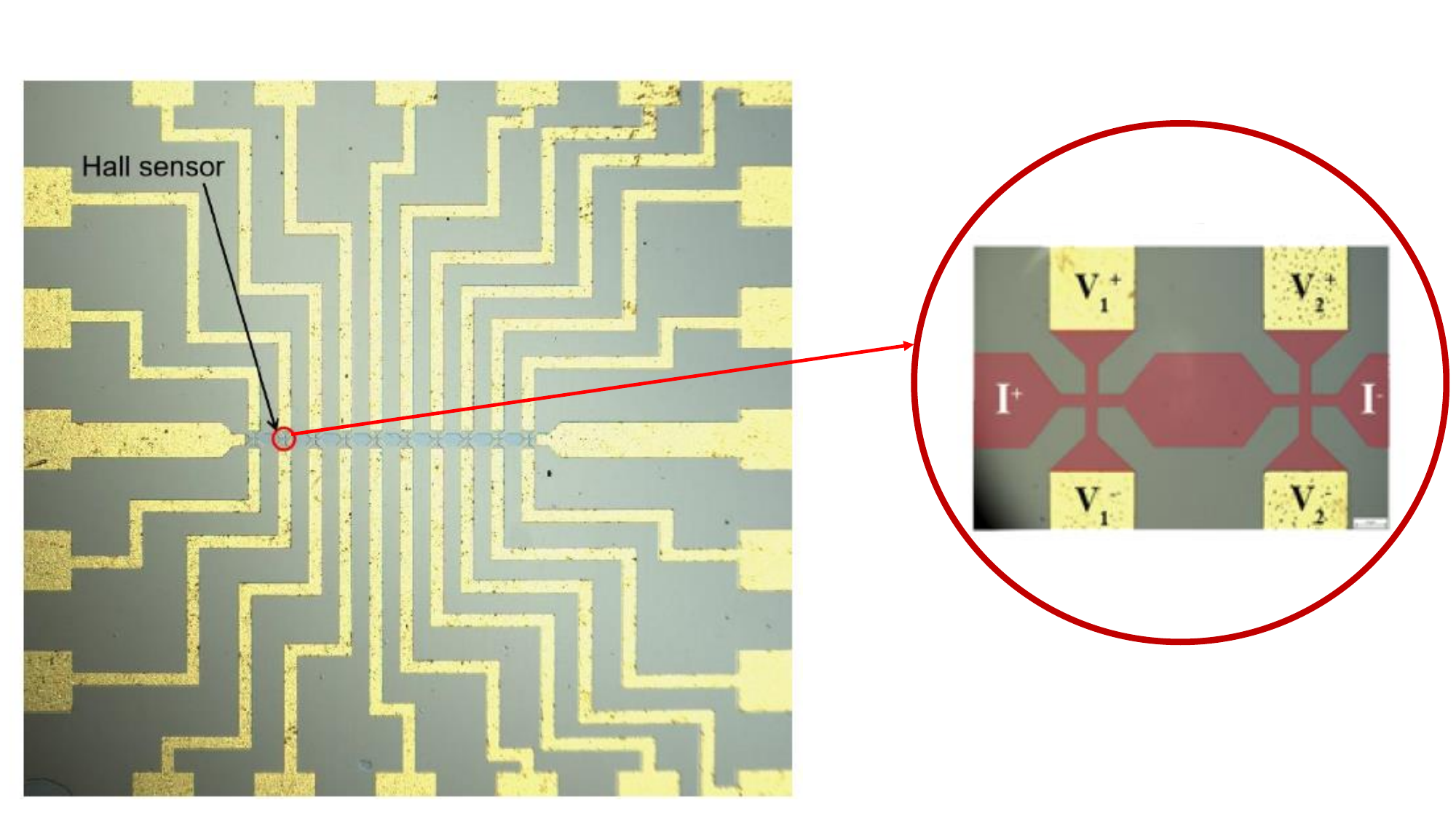}
\caption{\label{fig-SH}\justifying{\small{A picture of the array of miniature 2DEG Hall sensors, used to detect the local magnetic field perpendicular to their plane.}}}
\end{figure}

\end{document}